\documentclass[letter,11pt]{article}
\pdfoutput=1
\usepackage{heppub} 
\allowdisplaybreaks
\usepackage{hyperref}
\usepackage{graphicx}
\usepackage{amsmath}
\usepackage{amssymb} 
\usepackage[normalem]{ulem}
\usepackage{xspace}
\usepackage{slashed}
\usepackage[utf8]{inputenc}
\usepackage[T1]{fontenc}
\usepackage{subcaption}
\usepackage{scalerel,stackengine}
\usepackage[cal=cm,scr=euler]{mathalpha}

\newcommand{\ml}{\mathcal}

\newcommand{\T}{\mathcal{T}}

\preprint{CERN-TH-2025-008, MIT-CTP/5822}

\title{Dynamics of Heavy Quarks in Strongly Coupled $\mathcal{N}=4$ SYM Plasma} 

\author[1]{Krishna Rajagopal,}

\author[1,2]{Bruno Scheihing-Hitschfeld,}

\author[3]{and Urs Achim Wiedemann}

\affiliation[1]{
Center for Theoretical Physics, Massachusetts Institute of Technology, Cambridge, Massachusetts 02139, USA
}

\affiliation[2]{
Kavli Institute for Theoretical Physics, University of California, Santa Barbara, California 93106, USA
}

\affiliation[3]{Theoretical Physics Department, CERN, CH-1211 Gen\`eve 23, Switzerland}

\abstract{We calculate the probability distribution $P({\bf k})$ for a heavy quark with velocity $v$ propagating through strongly coupled $\mathcal{N}=4$ SYM plasma in the 't Hooft limit ($N_c\to \infty$, $\lambda=g^2 N_c \to \infty$) at a temperature $T$ to acquire a momentum ${\bf k}$ due to interactions with the plasma.
This distribution encodes the well-known drag coefficient $\eta_D$ and the transverse and longitudinal momentum diffusion coefficients $\kappa_T$ and $\kappa_L$. The jet quenching parameter $\hat{q}$ can be extracted from $P({\bf k})$ for $v = 1$. Going beyond these known Gaussian characteristics of $P({\bf k})$, our calculation determines all of the higher order and mixed moments to leading order in $1/\sqrt{\lambda}$ for the first time. These non-Gaussian features of $P({\bf k})$ include qualitatively novel correlations between longitudinal energy loss and transverse momentum broadening at nonzero $v$. We show that all higher moments scale characteristically with an effective temperature of the boosted plasma in the heavy quark rest frame, and we 
demonstrate that these non-Gaussian characteristics can be sizable in magnitude and even dominant in physically relevant situations.
We use these results to derive a Kolmogorov equation for the evolution of the probability distribution for the total momentum of a heavy quark that propagates through strongly coupled plasma. This evolution equation accounts for all higher order correlations between transverse momentum broadening and longitudinal energy loss, which we have calculated from first principles. It reduces to a Fokker-Planck equation when truncated to only include the effects of $\eta_D$, $\kappa_T$ and $\kappa_L$. Remarkably, while heavy quarks do not reach kinetic equilibrium with the plasma if evolved with this Fokker-Planck equation, by showing that the Boltzmann distribution is a static solution of the all-order Kolmogorov equation that we have derived we demonstrate that heavy quarks do reach kinetic equilibrium if evolved with this equation. Our results thus provide a dynamically complete framework for understanding the thermalization of a heavy quark that may be initially far from equilibrium in the strongly coupled $\mathcal{N}=4$ SYM 
plasma -- as well as new insight into heavy quark transport and equilibration in quark-gluon plasma.
}

\begin{document}

\maketitle

\section{Introduction}
Heavy quarks provide unique opportunities for understanding how single partonic degrees of freedom interact and exchange momentum and energy with, and are transported in, the hot plasmas of non-abelian gauge theories. This is so because heavy quarks can be viewed as long-lived test particles whose motion through the plasma is traceable and yet which interact with the plasma via the same forces as any quark constituent of the plasma. Indeed, in the QCD quark-gluon plasma (QGP) studied in ultra-relativistic heavy ion collisions, heavy flavor is conserved on electroweak time scales that are much longer than the lifetime of the produced plasma, and experimental techniques such as flavor tagging allow one to constrain and trace the propagation of individual heavy quarks through the plasma. When an experimentalist detects a heavy-flavor hadron in the final state of a heavy ion collision, they can know that the heavy quark therein was produced in an initial hard perturbatively calculable parton-parton scattering. From its production at time zero until the droplet of QGP in which it finds itself has cooled and frozen out more than 10 fm$/c$ later, this heavy quark interacted with, lost energy to, exchanged momentum with, and (if it had the time to lose enough of its initial momentum) ended up flowing along with and diffusing within the expanding droplet of QGP. All of these dynamics are together referred to as heavy quark transport. In the QCD QGP studied in these collisions, heavy quark transport is a non-perturbative phenomenon whose theoretical understanding requires strong coupling 
techniques.

In QCD, strong coupling techniques are scarce, and especially so for transport phenomena. A complementary approach towards understanding strongly coupled non-abelian plasmas emerged with the discovery of the gauge gravity duality~\cite{Maldacena:1997re}, combined with the early insight that at least some transport properties calculated for supersymmetric non-abelian quantum field theories with gravity duals are universal~\cite{Policastro:2001yc,Kovtun:2004de} and may thus be of direct relevance for understanding the QCD plasma produced in heavy-ion collisions. This triggered an avalanche of investigations of the plasma of ${\cal N}=4$ SYM theory in the ‘t Hooft strong coupling limit ($N_c \to \infty$, $\lambda \equiv g^2 N_c \to \infty$) and of related non-abelian plasmas. (For reviews, see Refs.~\cite{Casalderrey-Solana:2011dxg,DeWolfe:2013cua,Natsuume:2014sfa}.) In particular, for heavy quarks moving relative to the plasma with velocity $v$, the drag coefficient $\eta_D$ and the longitudinal and transverse momentum diffusion parameters $\kappa_L$ and $\kappa_T$ had been calculated early on in foundational works~\cite{Herzog:2006gh,Gubser:2006bz,Casalderrey-Solana:2006fio,Gubser:2006nz,Casalderrey-Solana:2007ahi}. All of these works calculated these parameters from a certain ``trailing string'' configuration, which describes the dynamics in the gravitational theory dual to a heavy quark in the field theory moving with velocity $v$ through strongly coupled plasma with temperature $T$.

These results came just after the dawn of the study of heavy quarks in heavy ion collisions using Langevin equations~\cite{vanHees:2005wb,Moore:2004tg}. The interest in obtaining them was great, as
Langevin dynamics describes the propagation of a heavy quark through a hot plasma in terms of precisely these three parameters ($\eta_D$, $\kappa_L$ and $\kappa_T$). When they are used as inputs for a Langevin description, the two parameters $\kappa_L$ and $\eta_D$ are (apparently) not independent as they are related via the Einstein relation
\begin{equation}
    \kappa_L = 2 T E\, \eta_D = 2 T M\, \eta_D \gamma\, ,
\label{eq:Einstein}
\end{equation}
where $T$ is the temperature of the plasma, $M$ is the mass of the heavy quark, and $E = \gamma M$ is its energy. This relation emerges in the Langevin approach as a consistency condition from the requirement that a heavy quark in a plasma with a constant temperature $T$ should eventually equilibrate, finding its way to a final state in which the probability distribution for its energy (and momentum) is a thermal Boltzmann distribution $\propto \exp(-E/T)$. This consideration does not constrain the velocity dependence of the transverse momentum diffusion constant $\kappa_T$, whose value in the strongly coupled ${\cal N}=4$ plasma, obtained via the AdS/CFT correspondence, is given by $\kappa_T=\pi\sqrt{\lambda}\gamma^{1/2}T^3$, a result that has been used as an input to many phenomenological investigations in QCD. 

From the beginning, however, a
critical issue with the trailing string results derived via the AdS/CFT correspondence, where $\eta_D = \tfrac{1}{2} \pi \sqrt{\lambda} T^2/M$ and $\kappa_L = \pi \sqrt{\lambda} \gamma^{5/2} T^3$, was identified. The dependence of $\kappa_L$ on the boost factor $\gamma$ was unexpected, as the calculated coefficients did not satisfy the Einstein relation. In the words of Steven Gubser~\cite{Gubser:2006nz}:

\begin{quote}
{\it ``Because of the dramatic failure of the Einstein relation, it doesn’t make sense to plug the trailing string predictions for $\eta_D$, $\kappa_L$ and $\kappa_T$ into a Langevin description of a finite mass quark: it won’t equilibrate to a Maxwell-Boltzmann distribution, due to the largeness of $\kappa_L$ as compared to $\eta_D$ at highly relativistic speeds. Perhaps what is needed is a stochastic treatment of the quasi-normal modes of the trailing string, ...''}
\end{quote}

\noindent
Many subsequent investigations~\cite{deBoer:2008gu,Son:2009vu,Giecold:2009cg,Casalderrey-Solana:2009ifi,Caron-Huot:2011vtx} considered a stochastic treatment of the fluctuations on the string to construct a Langevin description of the motion of a heavy quark with drag coefficient $\eta_D$, yet upon using the result for $\kappa_L$ equilibration to a Boltzmann distribution remained elusive.

Although the puzzle presented by the $\gamma$-dependence of $\kappa_L$ has cast doubt on the interpretation of these results since they were obtained, this has not stopped many authors from making many successful qualitative and semi-quantitative uses of the AdS/CFT results for $\eta_D$ and $\kappa_T$ in the phenomenology of heavy-flavor transport. Further confidence in these comparisons comes from current lattice  QCD calculations of $T^3/\kappa$ (where $\kappa$ is the $v\rightarrow 0$ limit of both $\kappa_T$ and $\kappa_L$) which show that in the strongly coupled QGP of QCD at temperatures up to 300 MeV 
this quantity is roughly as small as the AdS/CFT result indicates~\cite{Altenkort:2023eav}.

In the present work, we shall revisit the problem of describing the evolution of the momentum of a heavy quark in the strongly coupled ${\cal N}=4$ SYM plasma. Instead of focusing only on the quantities described above, we calculate the full probability $P({\bf k})$ that the momentum of the heavy quark changes over a time period $\T$ by an arbitrary momentum ${\bf k}$ that may include transverse or longitudinal components or both. 
$P({\bf k})$ describes drag and energy loss (via a decrease in the mean longitudinal momentum) as well as fluctuations in and correlations among longitudinal and transverse momentum.
It can be related to the thermal expectation value of a certain Wilson loop using Heavy Quark Effective Theory (HQET), in which the interactions of a heavy quark with the plasma at leading order in $1/M$ are described by a Wilson line that follows the heavy quark trajectory, characterized by its velocity $v$. 
Multiplying the amplitude to acquire a momentum ${\bf k}$ by its complex conjugate and averaging over the plasma degrees of freedom leads to 
\begin{align}
    P({\bf k}) &\propto \int d^3 {\bf L} \, e^{ - i  {\bf k} \cdot {\bf L} } \langle W[\mathcal{C}] \rangle_T({\bf L}) \, , \label{eq:P-W-intro}
\end{align}
where the Wilson Loop contour $\mathcal{C}$, depicted in Section~\ref{sec:ads-cft}, includes two long straight segments with spacetime slope $v$ and spatial separation ${\bf L}$.
AdS/CFT studies of this object have previously been formulated both from HQET~\cite{Casalderrey-Solana:2007ahi} for $0 < v < 1$ and from Soft-Collinear Effective Theory (SCET)~\cite{DEramo:2010wup} for $v = 1$, although with an emphasis on transverse momentum broadening.

\begin{figure}
    \centering
    \includegraphics[width=0.88\textwidth]{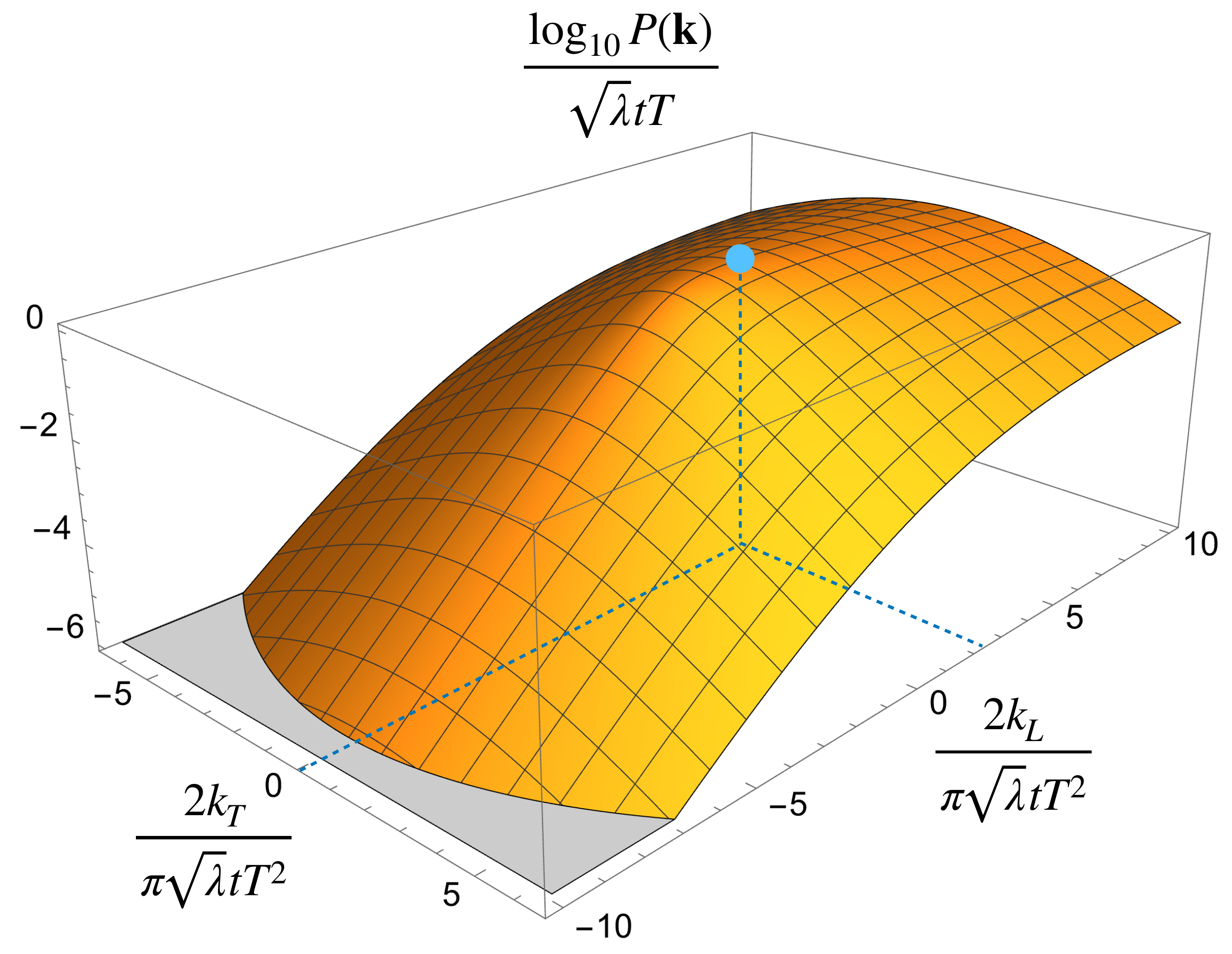}
    \caption{The probability distribution we calculate in this work, for a heavy quark moving for a time $t$ with a speed $v=0.9$ relative to the rest frame of the strongly coupled plasma with temperature $T$ and 't Hooft coupling $\lambda = g^2 N_c$. The momentum ${\bf k}$, with longitudinal and transverse components $k_L$ and $k_T$, is the momentum 
    transferred from the heavy quark to the plasma. That is, positive $k_L$ corresponds to momentum lost by the heavy quark. The values of $k_L$ and $k_T$ reached after time $t$ are to be sampled from the probability distribution $P({\bf k})$ depicted in orange. The blue dot shows the position of the maximum of the distribution, given by $k_L = \pi \sqrt{\lambda} t T^2 \gamma v/2$, which determines the heavy quark drag coefficient first calculated in Refs.~\cite{Herzog:2006gh,Gubser:2006bz}. The curvatures of the orange surface about the blue dot in the longitudinal and transverse directions determine $\kappa_L$ and $\kappa_T$, first calculated in Refs.~\cite{Casalderrey-Solana:2006fio,Gubser:2006nz,Casalderrey-Solana:2007ahi}. }
    \label{fig:P-of-k-intro}
\end{figure}

Beginning from the results of Refs.~\cite{Skenderis:2008dg,Skenderis:2008dh}, we shall set up and carry out a complete calculation of $P({\bf k})$ for an arbitrary relative velocity $v$ between the rest frames of the heavy quark and the plasma.
In Section~\ref{sec:ads-cft}, we explain the holographic setup of this calculation in detail. In Section~\ref{sec:momentum-broadening}, we then derive and evaluate an explicit analytic expression for the momentum transfer distribution $P({\bf k} )$. We recover the known results for $\eta_D$, $\kappa_L$ and $\kappa_T$ from the  position of the maximum of $P({\bf k})$ and the Gaussian curvatures about the maximum. Although (as we shall review in Section~\ref{sec:regime}) 
the probability distribution $P({\bf k})$ only describes heavy quark transport for $\sqrt{\gamma}\ll M/(\sqrt{\lambda}T)$ and therefore not for $v\rightarrow 1$ at any finite value of $M$, we shall show that we can nevertheless
also rederive the known 
result~\cite{Liu:2006ug,Liu:2006he,DEramo:2010wup} for the jet quenching parameter $\hat{q}$ in strongly coupled plasma 
from the curvature of $P(k_T,k_L=0)$ about $k_T=0$ at $v=1$. As one may readily see from Figure~\ref{fig:P-of-k-intro}, our approach will not be limited to reproducing these known Gaussian characteristics. Rather, our calculation gives explicit access to the entire velocity-dependent $P({\bf k})$ at leading order in $1/\sqrt{\lambda}$. This allows us in Section~\ref{sec:non-gaussian} to provide explicit expressions for {\it all} of the previously unexplored non-Gaussian higher moments of $P({\bf k})$, as well as all of the correlations (sometimes called mixed moments) between the longitudinal and transverse momenta exchanged between the heavy quark and the strongly coupled plasma. We shall find a characteristic scaling of the higher order and mixed moments and discuss the qualitatively novel correlations between longitudinal and transverse dynamics which the mixed moments describe. 

In Section~\ref{sec:consequences}, we exploit our complete knowledge of $P({\bf k})$ to derive an evolution equation for the probability distribution of the heavy quark momentum ${\bf p}$ (not just its small change ${\bf k}$ conditioned on a near-constant velocity $v$), which we will refer to as a Kolmogorov equation. 
If restricted to the Gaussian characteristics of $P({\bf k})$, this Kolmogorov evolution equation reduces to a Fokker-Planck equation with parameters $\eta_D$, $\kappa_L$ and $\kappa_T$ --- violating the Einstein relation. The restriction to Gaussian characteristics thus exhibits the familiar thermalization problem highlighted by Gubser. In contrast, we are now able to argue in 
Section~\ref{sec:equil-distr}, that the full Kolmogorov equation which we have derived describes dynamics according to which heavy quarks, even if initially propagating at a highly relativistic speed, {\it will} eventually equilibrate to a Boltzmann distribution. 
We show this by demonstrating that a Boltzmann distribution for the momentum of the heavy quark is a stationary solution to the full Kolmogorov equation to leading order in $T/M$, a demonstration that depends upon all of the higher moments of $P({\bf k})$ that we calculated in Section~\ref{sec:momentum-broadening} and so is impossible to realize via the Fokker-Planck truncation.
We thus realize the first complete description of heavy quark transport in strongly coupled plasma, including momentum loss and momentum broadening followed by stopping, equilibration and diffusion.

Since we have found an evolution equation that can describe the entire thermalization process of a heavy quark in the strongly coupled plasma of ${\cal N}=4$ SYM, we hope that this result can be the basis for future studies of heavy quark thermalization in the QGP of QCD and that it proves useful for future phenomenological discussions of heavy quark transport in heavy ion collisions. In Section~\ref{sec:conclusions}, we therefore look ahead and comment in some detail on these perspectives.

\section{The AdS/CFT Setup} \label{sec:ads-cft}

The prescription to calculate the expectation value of Wilson loops in $\mathcal{N}=4$ supersymmetric Yang-Mills (SYM) theory at large $N_c$ and strong coupling using the AdS/CFT correspondence dates back to the works by Maldacena~\cite{Maldacena:1998im} and Rey \& Yee~\cite{Rey:1998ik}. We begin in Section~\ref{sec:W-setup} by discussing this setup in Euclidean signature, describe how to extend it to real time using the Schwinger-Keldysh contour in Section~\ref{sec:SK}, which requires introducing a complex path in the time coordinate to access real-time dynamics at finite temperature, and only later in Section~\ref{sec:Wloop-SK} we specialize to our Wilson loop of interest on the Schwinger-Keldysh contour. In Section~\ref{sec:Wloop-Prob} we discuss how the AdS/CFT setup gives us direct access to the momentum transfer distribution $P({\bf k})$ for a hard particle propagating through strongly coupled plasma.

\subsection{Wilson Loops in AdS/CFT} \label{sec:W-setup}

The expectation value of a Wilson loop along a contour $\mathcal{C}$ in a strongly coupled quantum field theory, specifically in $\mathcal{N}=4$ SYM in the limits $N_c \to \infty$, $\lambda = N_c g^2 \to \infty$, can be calculated by solving the classical equations of motion for a string in a 10-dimensional ${\rm AdS}_5 \times S_5$ space~\cite{Rey:1998ik,Maldacena:1998im,Drukker:1999zq}:
\begin{align} \label{eq:duality}
\left\langle W_{S}[\mathcal{C};\hat{n}] \right \rangle = \exp \left\{ -  (\mathcal{S}_{\rm NG}[\Sigma(\mathcal{C};\hat{n})] - \mathcal{S}_{0})  \right\} \, .
\end{align}
Here, the Wilson loop operator is written as a path-ordered exponential 
\begin{align} \label{eq:W-loop-S}
W_{S}[\mathcal{C};\hat{n}] = \frac{1}{N_c} {\rm Tr}_{\rm color} \! \left[ \mathcal{P} \exp \left( ig \oint_{\ml{C}} ds \, T^a \left[  A^a_\mu \, \dot{x}^\mu + \hat{n}(s) \cdot {\vec{\phi}}^a \sqrt{\dot{x}^2} \right] \right) \right] \, ,
\end{align}
where $\dot{x}^\mu(s) = dx^\mu(s)/ds$ parametrizes the path $\mathcal{C}$ over which the Wilson loop goes in Minkowski coordinates. The operators in the path-ordered exponential are the gauge field $A_\mu$ and the six Lorentz scalar fields $\vec{\phi} = (\phi^1, \ldots, \phi^6)$ in the adjoint representation of SU($N_c$) that are part of the matter content of $\mathcal{N}=4$ SYM. These scalars enter the Wilson loop coupled to a direction $\hat{n}(s) \in S_5$ that specifies the direction along which the string (to be introduced in the next paragraph) is oriented in the $S_5$ inside the 10-dimensional ${\rm AdS}_5 \times S_5$ space.

The right hand side of~\eqref{eq:duality} is specified by the Nambu-Goto action of a string in ${\rm AdS}_5 \times S_5$
\begin{align} \label{eq:NG-action}
\mathcal{S}_{\rm NG}[\Sigma] = \frac{1}{2\pi \alpha'} \int d\sigma \, d\tau \sqrt{\det \left( g_{\mu \nu} \partial_\alpha X^\mu \partial_\beta X^\nu \right) } \, .
\end{align}
This is a functional of a string configuration $\Sigma$ described by $X^\mu(\tau,\sigma) \in {\rm AdS}_5 \times S_5$, with $\mu \in \{0,1,\ldots,9\}$\@. In Eq.~\eqref{eq:duality}, the action is evaluated on the extremal surface configuration $\Sigma(\mathcal{C}; \hat{n})$ (also referred to as the string worldsheet) that minimizes the Nambu-Goto action $\mathcal{S}_{\rm NG}$ subject to Dirichlet boundary conditions given by $\mathcal{C}$ and $\hat{n}$ at the asymptotic boundary of AdS${}_5$\@. The action $\mathcal{S}_{\rm NG}$  is obtained by carrying out the path integral over the degrees of freedom of the surface via the saddle point approximation by virtue of $\lambda \to \infty$ (note that $\lambda$ is related to $\alpha'$ by $\sqrt{\lambda} = R^2/\alpha'$. $R$ is the AdS radius, to be introduced below.). The subtraction of $\mathcal{S}_{0}$ in Eq.~\eqref{eq:duality} removes the energy associated with the (formally infinite) mass $M$ of the heavy quark propagating along $\mathcal{C}$ from the Nambu-Goto action. It is given by $\mathcal{S}_{0} = M L[\mathcal{C}]$, where $L[\mathcal{C}] = \int ds \sqrt{\dot{x}_\mu \dot x^{\mu} }$ is the magnitude of the invariant spacetime interval that the path $\mathcal{C}$ traverses. To carry out this subtraction one needs to introduce a regulator into the calculation, which may only be removed after the subtraction has taken place.

This setup allows one to calculate expectation values of the above Wilson loops in $\mathcal{N}=4$ SYM in the canonical ensemble at any temperature~\cite{Witten:1998qj}. The temperature $T = 1/\beta$ enters this setup as a parameter of the (asymptotically) ${\rm AdS}_5 \times S_5$ metric
\begin{equation}
    ds^2 = \frac{R^2}{z^2} \left[ f(z) \, d\tau^2 + d{\bf x}^2 + \frac{dz^2}{f(z)} + z^2 d\Omega_5^2 \right] \, , \label{eq:AdS-T-metric}
\end{equation}
where $\tau \in (0,\beta)$ is the periodic Euclidean time coordinate of thermal field theory~\cite{Bellac:2011kqa,Laine:2016hma}, $z$ is the radial ${\rm AdS}_5$ coordinate, and
\begin{equation}
    f(z) = 1 - (\pi T z)^4 \, .
\end{equation}
This metric describes a black brane inside an asymptotically ${\rm AdS}_5 \times S_5$ spacetime, whose position sets the temperature of the dual field theory.

As shown by Alday \& Maldacena~\cite{Alday:2007he} and Polchinksi \& Sully~\cite{Polchinski:2011im}, the usual gauge theory Wilson loop
\begin{equation}
    W[\mathcal{C}] = \frac{1}{N_c} {\rm Tr}_{\rm color} \! \left[ \mathcal{P} \exp \left( ig \oint_{\ml{C}} ds \, T^a \left[  A^a_\mu \, \dot{x}^\mu \right] \right) \right] \label{eq:W-loop}
\end{equation}
is obtained by integrating~\eqref{eq:duality} over $\hat{n}$. This integration leads to imposing Neumann boundary conditions along the $S_5$ directions on the worldsheet~\cite{Alday:2007he,Polchinski:2011im} (see also~\cite{Drukker:1999zq} for earlier work). For the problem discussed in the present paper, integration over  $\hat{n}$ at leading order in the strong coupling limit $\lambda \to \infty$ will select a single, fixed value $\hat{n}_0$ that does not affect the final result (see Section~\ref{sec:SK-field} for more details). We note as an aside that in the study of correlation functions for quarkonium transport~\cite{Nijs:2023dbc,Nijs:2023dks}, $\hat{n}$ plays a relevant, in fact an important, role.

\subsection{The Schwinger-Keldysh 
Contour} \label{sec:SK}

Before discussing how to calculate Eq.~\eqref{eq:P-W-intro} using the AdS/CFT correspondence, it is instructive to discuss its properties on the quantum field theory side of the duality. We do this in Section~\ref{sec:SK-field} and then proceed to describe its holographic realization in Section~\ref{sec:SK-holo}.

\subsubsection{The Schwinger-Keldysh Contour in Field Theory} \label{sec:SK-field}

\begin{figure}[t]
    \centering
    \includegraphics[width=\linewidth]{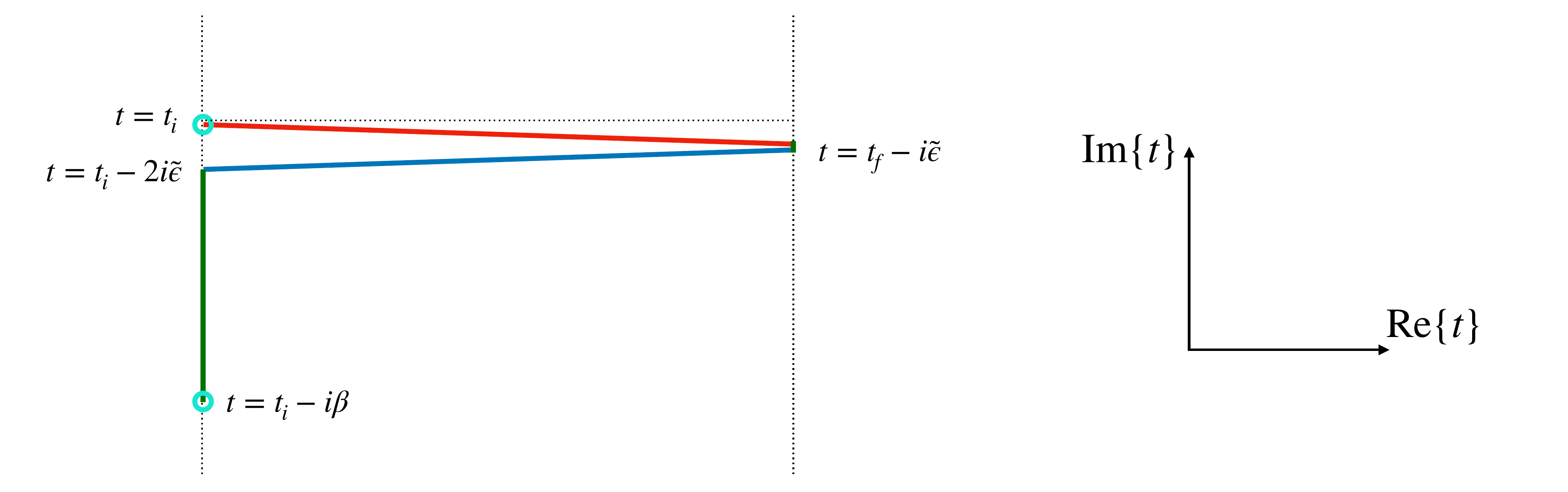}
    \caption{The Schwinger-Keldysh contour. In red, the time-ordered branch. In blue, the anti-time-ordered branch. In terms of the quantities introduced in the main text, we have denoted $\tilde{\epsilon} = \epsilon (t_f -t_i)$ for brevity.}
    \label{fig:SK-contour}
\end{figure}

Many observables in quantum field theory can be related to time-ordered correlation functions, but there are some that cannot be formulated in terms of a path integral of fields over a single Minkowski spacetime manifold. The Wilson loop in Eq.~\eqref{eq:P-W-intro} is an example of this, since the two long, antiparallel Wilson lines that comprise it have different operator orderings~\cite{Casalderrey-Solana:2006fio,Casalderrey-Solana:2007ahi,DEramo:2010wup}. It is part of a broad class of real-time operator expectation values in a thermal ensemble that can be formulated along a complex time path given by the so-called Schwinger-Keldysh (SK) contour drawn in Figure~\ref{fig:SK-contour}.

As first noted in Refs.~\cite{Casalderrey-Solana:2006fio,Casalderrey-Solana:2007ahi}, the desired operator ordering of the Wilson loop is enforced by having it wind around the SK contour, with each segment having a unique interpretation in terms of the building blocks of~\eqref{eq:P-W-intro}:
\begin{enumerate}
    \item The segment on the time-ordered part of the SK contour corresponds to the scattering amplitude. Its complex time path is parametrized by $t_i + (1-i\epsilon)t$, with $t \in (t_i,t_f)$.
    \item The segment on the 
    anti-time-ordered part of the SK contour corresponds to the conjugate of the scattering amplitude. Its complex time path is parametrized by $t_f - i \epsilon(t_f-t_i) - (1 +i\epsilon)(t_f - t)$, with $t \in (t_i,t_f)$.
    \item the segment on the imaginary time part of the SK contour corresponds to the preparation of the plasma's thermal ensemble in the presence of a point color charge. Its complex time path is parametrized by $t_i - i\tau_E$, with $\tau_E \in (2\epsilon(t_f-t_i), \beta)$.
\end{enumerate}

\begin{figure}[t]
    \centering
    \includegraphics[width=0.99\linewidth]{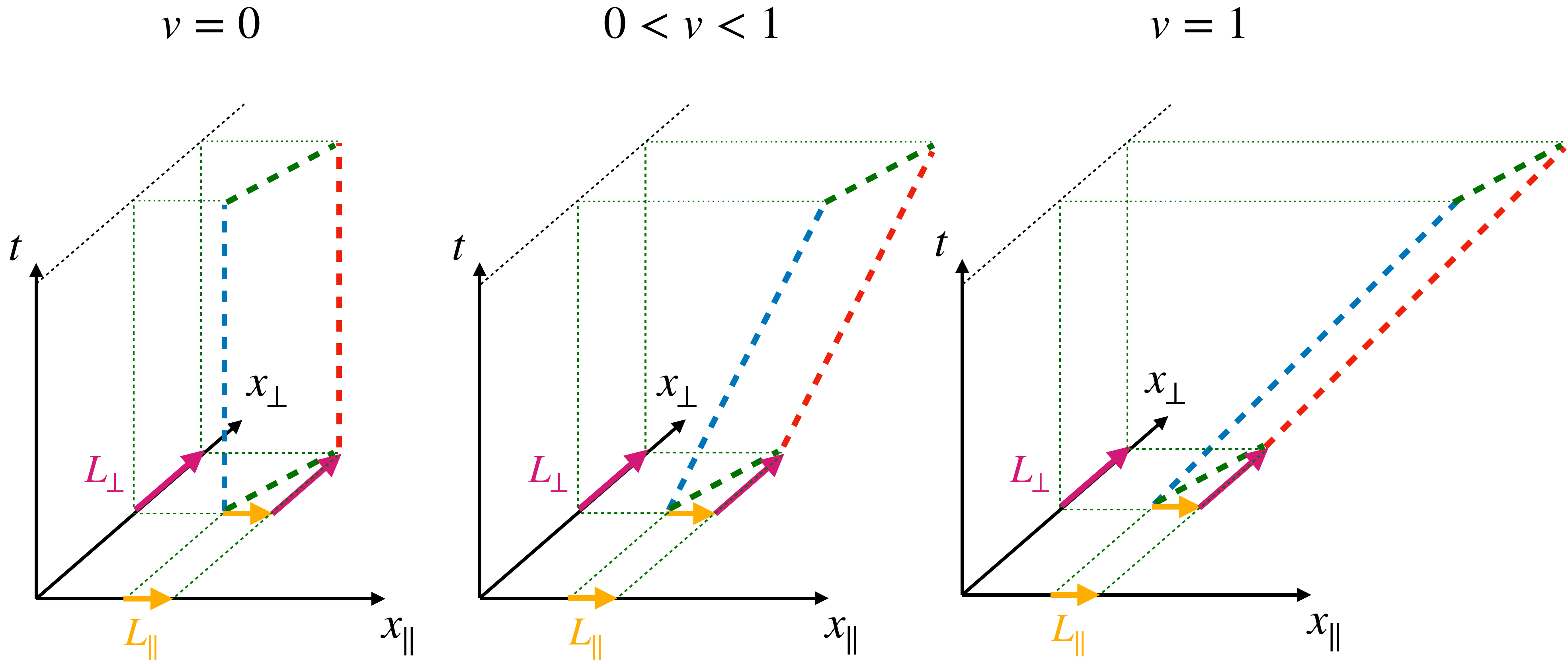}
    \caption{The family of Wilson loops of interest in this work. The thick dashed lines describe the path $\mathcal{C}$ that defines each Wilson loop. We use the same colors as in Figure~\ref{fig:SK-contour} to reflect the fact that the two antiparallel lines live on the corresponding branches of the SK contour, and only meet at the initial and final times ($-\T/2$ and $\T/2$, respectively). The slope of the long dashed lines in spacetime is determined by $v$. Their relative position in space is specified by $L_\perp$ and $L_\parallel$, depicted by the red and orange vectors. After evaluating the expectation value of the Wilson loop, these quantities will be integrated over in order to determine $P({\bf k})$, see Eq.~(\ref{eq:P-W-intro}).} 
    \label{fig:WLoop-diagram}
\end{figure}

In the following, we are interested in Wilson loops  that involve both a velocity $v$ of the trajectory relative to the plasma's rest frame, and a separation in their spatial positions,  see Figure~\ref{fig:WLoop-diagram}. Gauge invariance requires then that the two long, antiparallel Wilson lines be connected at $t = \T/2 \to +\infty$ by a transverse gauge link going over a spacelike path. How this additional Wilson line is arranged in space is of no concern to us, as it is not dependent on the length of the long antiparallel Wilson lines and it thus will not affect the calculation of the leading behavior in the large $\T$ limit of $P({\bf k})$. Furthermore, since this gauge link at $t = \T/2 \to +\infty$ is spacelike, no operator ordering issues appear and it may thus be inserted at the end of the time-ordered branch or at the end of the anti time-ordered branch of the SK contour. Also at $t = -\T/2 \to -\infty$, the details of how the Wilson line is arranged in space will not affect the calculation of the leading behavior in the large $\T$ limit of $P({\bf k})$.\footnote{We comment that the operator ordering aspects can be more complicated at $t = -\T/2 \to -\infty$, because the precise way in which the Wilson line is arranged along the imaginary time segment of the SK contour will reflect different initial state preparations.}

We now return to the dependence of the Wilson loop~\eqref{eq:W-loop-S} on $\hat{n}$. Physically, we require that in the limit of vanishing spatial separation, the two long Wilson lines become the exact inverse of each other. For the $A_\mu \dot{x}^\mu$ term in the path-ordered exponential, this is achieved due to $\dot{x}$ flipping sign. For the $\hat{n} \cdot \phi \sqrt{\dot{x}^2}$ term, a constant $\hat{n}$ achieves this because the analytic continuation of $\sqrt{\dot{x}^2}$ from Euclidean signature onto each of the two real-time segments of the SK contour will yield an overall sign of difference as a consequence of the different $\epsilon$ prescriptions. It is thus the properties of the SK contour that make this the appropriate configuration of $\hat{n}$ to study.\footnote{This is to be contrasted with the case of correlation functions relevant for quarkonium transport~\cite{Nijs:2023dbc,Nijs:2023dks}, where antipodal points on the $S_5$ are the appropriate choice for the Wilson lines to be the exact inverse of each other. This is because, in that calculation unlike in this one, the real-time segments of the relevant Wilson loop are on the same branch of the SK contour, and so no sign flip in $\sqrt{\dot{x}^2}$ takes place.}
This is consistent with using the prescription~\cite{Alday:2007he,Polchinski:2011im} of integrating~\eqref{eq:W-loop-S} over $\hat{n}$ to calculate the pure gauge Wilson loop~\eqref{eq:W-loop}, which automatically satisfies the requirement that the antiparallel Wilson lines are the inverse of each other when their spatial separation is zero. Therefore, the expectation value of the Wilson loop we calculate in what follows is in fact the thermal expectation value of the pure gauge Wilson loop~\eqref{eq:W-loop}, which is the object that appears in the definition of $P({\bf k})$ in Eq.~\eqref{eq:P-W-intro}.

\subsubsection{The Holographic Realization of the Schwinger-Keldysh Contour} \label{sec:SK-holo}

While real-time calculations using the AdS/CFT correspondence at a finite temperature have a long history, with perhaps the most famous examples being that of the shear viscosity of strongly coupled $\mathcal{N}=4$ plasma~\cite{Policastro:2001yc}, the heavy quark drag coefficient~\cite{Gubser:2006bz,Herzog:2006gh}, the heavy quark diffusion 
constants~\cite{Casalderrey-Solana:2006fio,Casalderrey-Solana:2007ahi,Gubser:2006nz}, the jet quenching parameter~\cite{Liu:2006ug,DEramo:2010wup}, and the heavy quark potential in a ``hot wind''~\cite{Liu:2006he,Liu:2006nn,Chernicoff:2006hi},
the question of how to do field theory calculations on the SK contour using holographic methods was first addressed comprehensively in works by Skenderis and van Rees~\cite{Skenderis:2008dg,Skenderis:2008dh}. We now review the ingredients of their work that we will use in what follows. We note that further developments have taken place since then~\cite{Glorioso:2018mmw}, allowing for simpler calculations of $n$-point correlation functions of local operators. It would be interesting to investigate extending these methods to the calculation of Wilson lines.

The main result of Refs.~\cite{Skenderis:2008dg,Skenderis:2008dh} is that the holographic dual to a field theory path integral on the SK contour (or any closed time path) is given by a gravitational theory on a manifold whose asymptotic boundary is equivalent to the SK contour. For each segment of the SK contour there is a corresponding patch of the manifold that describes the bulk of the AdS space, which is spanned by an extension of the time coordinate $t$ that goes over each of the field theory SK segments, and whose boundary in the radial AdS direction $z \to 0$ is exactly the corresponding field theory SK contour segment. At the temporal ends of each of these patches dual to each segment of the field theory SK contour, appropriate matching conditions between the fields must be imposed, analogously to the matching conditions that are imposed on the field theory path integral. Concretely, the values of the fields and the canonical momenta have to be matched to fully specify the gravitational theory.

To be explicit, the holographic realization of the SK path integral for $\mathcal{N}=4$ SYM that we will use here involves three patches, each of them given by the metric~\eqref{eq:AdS-T-metric} with the substitutions:
\begin{enumerate}
    \item $\tau \to i t_i + (i + \epsilon)(t-t_i)$ for the time-ordered branch of the SK contour. $t$ goes from $t_i$ to $t_f$, where matching conditions are imposed with $t = t_f$ on the anti-time-ordered branch.
    \item $\tau \to it_f + \epsilon(t_f-t_i) + (-i + \epsilon)(t_f - t)$ for the anti-time-ordered branch of the SK contour. $t$ goes from $t_f$ to $t_i$, where matching conditions are imposed with $\tau = 2\epsilon (t_f-t_i)$ on the imaginary time segment.
    \item $\tau \to i t_i + \tau_E$ for the imaginary time segment of the SK contour. $\tau_E$ goes from $2\epsilon (t_f-t_i)$ to $ i t_i + \beta$, where matching boundary conditions are imposed with $t = t_i$ on the time-ordered branch.
\end{enumerate}
We note that this construction has been described previously in Refs.~\cite{vanRees:2009rw,DEramo:2010wup}. For a pedagogical introduction, see Section 5 of~\cite{Haehl:2024pqu}. We refer the reader to those works for geometric visualizations of this holographic construction.

One important consequence of the above is that the black brane horizon, $z = 1/(\pi T)$, is common to all three patches. This is so since the proper length spanned by the $t$ coordinate vanishes at this location. Therefore, all points with $z = 1/(\pi T)$ and differing values of the $t$ coordinate are actually the same point, a familiar result from the Euclidean Schwarzschild metric arising here for the Euclidean AdS black brane. As such, the values of the fields approaching the horizon from the different patches dual to each segment of the SK contour satisfy an additional matching condition, which we now introduce for the Wilson loop of our interest.

\subsection{Calculating Wilson Loops with the Holographic Schwinger-Keldysh Path Integral} \label{sec:Wloop-SK}

We start by writing the right-hand side of Eq.~\eqref{eq:duality} in the form of a path integral
\begin{equation}
    \langle W[\mathcal{C}] \rangle_T = \int D X_R D X_L D X_E \exp \left( - S[X_R] - S[X_L] - S[X_E] \right) \, .
\end{equation}
Here, $X_R$ lives on the time-ordered branch of the SK contour, $X_L$ lives on the anti-time-ordered branch of the SK contour, and $X_E$ lives on the Euclidean branch,
as discussed in Section~\ref{sec:SK-field}. The three contributions satisfy the matching conditions discussed in Section~\ref{sec:SK-holo}. 
The subscript $T$ indicates that the average is taken over a thermal ensemble.

We are interested in the large $\T$ limit of $\langle W[\mathcal{C}] \rangle_T$. Since $S[X_E]$ is non-extensive in $\T$, the result for $\langle W[\mathcal{C}] \rangle_T$ becomes independent of $S[X_E]$ in the large $\T$ limit and simplifies in this limit to
\begin{equation}
    \langle W[\mathcal{C}] \rangle_T = \int D X_R D X_L \exp \left( - S[X_R] - S[X_L] \right) \, .
\end{equation}
At intermediate times $t$, far away from the endpoints of the Wilson loop $- \T/2 \ll t \ll \T/2$, and assuming that $\T T \gg 1$, the worldsheet $X = (t,{\bf x},z)$ can be parametrized in terms of a time variable $\tau$ and the radial AdS coordinate $z$, both real, as
\begin{align}
    X_R(\tau, z) &= \big( \tau (1 - i\epsilon) , x_{R1}(z), x_{R2}(z) , v\tau + x_{R3}(z), z \big) \, , \\
    X_L(\tau, z) &= \big( \tau (1 + i\epsilon) , x_{L1}(z), x_{L2}(z) , v\tau + x_{L3}(z), z \big) \, ,
\end{align}
where the $i\epsilon$'s reflect the fact that the tangent vector to the SK contour always has a small imaginary part in the $t$ coordinate, consistent with our discussion in the previous 
Sections. There is no $\tau$ dependence in any of the spatial coordinates of the worldsheet because of the time-translation symmetry that emerges in the limit $\T T \gg 1$. Since this symmetry emerges in both the action and the boundary conditions, the string configurations that give the dominant contributions to the path integral also inherit this symmetry. We have omitted the $\hat{n} \in S_5$ coordinates because, as argued before, they are set to a constant value and do not play any further role in the subsequent calculation. Direct substitution into the Nambu-Goto action~\eqref{eq:NG-action} then shows that
\begin{align}
    S[X_R] &=  \frac{\sqrt{\lambda} \T }{2\pi} \int_0^{z_h} \frac{dz}{z^2} \sqrt{ - \left( f_-(z) x_{R3}'^2 + (f_-(z) - v^2) (x_{R1}'^2 + x_{R2}'^2) + \frac{f_-(z)-v^2}{f} \right) } \, , \label{eq:SXR} \\
    S[X_L] &=  \frac{\sqrt{\lambda} \T }{2\pi} \int_0^{z_h} \frac{dz}{z^2} \sqrt{ - \left( f_+(z) x_{L3}'^2 + (f_+(z) - v^2) (x_{L1}'^2 + x_{L2}'^2 ) + \frac{f_+(z)-v^2}{f} \right) } \, . \label{eq:SXL}
\end{align}
Here, $\T$ is the extent of the loop along the $t$ coordinate, $z_h = 1/(\pi T)$, $'$ denotes $d/dz$, and we have defined 
\begin{equation}
    f_\pm(z) \equiv (1 \pm i\epsilon) f(z) \, . \label{eq:fpm}
\end{equation}

It remains to implement the matching conditions between $X_R$ and $X_L$, and the boundary conditions for $x_{Ri}$ and $x_{Li}$ at $z=0$. By specializing to the region $-\T/2 \ll t \ll \T/2$, we have implicitly assumed that the string configuration does not depend on the details of how the worldsheet is closed at $t = \pm \T/2$. There is then only one condition left to impose: continuity of the worldsheet at the horizon $z = z_h = 1/(\pi T)$. We may enforce this by inserting an explicit Dirac delta-function in the path integral for the Wilson loop:
\begin{align}
    \langle W[\mathcal{C}] \rangle_T &= \int D x_{Rj} D x_{Lj} \exp \left( - S[X_R] - S[X_L] \right) \delta^{(3)}(x_{Rj}(z_h) - x_{Lj}(z_h) ) \nonumber \\
    &= \int D x_{Rj} D x_{Lj} dc_j \exp \left( - S[X_R] - S[X_L] - i c_j \left[ x_{Lj}(z_h) - x_{Rj}(z_h) \right]  \right)  \, . \label{eq:W-loop-bdy-cond-hor}
\end{align}
For the following, it is useful to rewrite the expression~\eqref{eq:W-loop-bdy-cond-hor} in terms of integrations over $x_{Rj}'$ and $x_{Lj}'$ and in terms of a rescaled Lagrange multiplier 
$C_i \equiv \frac{2}{\sqrt{\lambda}T \T}c_i$:
\begin{equation}
    \langle W[\mathcal{C}] \rangle_T = \int D x_{Rj}' D x_{Lj}' dC_i \exp \left( - \sqrt{\lambda} T \mathcal{T} S_{R+L+C} \right) \,  \label{eq:W-loop-Stot}
\end{equation}
with 
\begin{equation}
    \sqrt{\lambda} T \mathcal{T} S_{R+L+C} = S[X_R] + S[X_L] - i \frac{\sqrt{\lambda} T \T}{2 } C_j \left(  \int_0^{z_h} dz \, x_{Lj}' - \int_0^{z_h} dz \, x_{Rj}' + L_j \right) \, , \label{eq:S-tot}
\end{equation}
where $L_j$ is the $j$'th component of ${\bf L}$, the spatial separation of the two antiparallel Wilson lines. We have used the boundary conditions at $z=0$ to rewrite the  difference $i c_j \left[x_{Lj}(z_h) - x_{Rj}(z_h)\right]$ in \eqref{eq:W-loop-bdy-cond-hor} as the last term of \eqref{eq:S-tot}. Changing integration variables affects the prefactor outside the exponential in the path integral~\eqref{eq:W-loop-bdy-cond-hor}. However, this is unimportant for the following, since we work in the large $\lambda$ limit and do not keep track of this prefactor, which only affects the result at subleading order in $1/\sqrt{\lambda}$.

\subsection{Relating the Wilson Loop to the Momentum Transfer Distribution} \label{sec:Wloop-Prob}

As we show now, 
although we realized this only after spending considerable effort on explicit calculation of the thermal expectation value of the Wilson loop,
it is in fact easier to calculate $P({\bf k})$ than it is to calculate $\langle W[\mathcal{C}] \rangle_T$. To this end, we define $\tilde{S}_{\rm tot}$ as the result of carrying out the functional integral over $x_{Rj}'$ and $ x_{Lj}'$ in Eq.~\eqref{eq:W-loop-Stot}, without the piece that depends on $L_j$:
\begin{equation}
    \exp \left(-\sqrt{\lambda} T \T \left[ \tilde{S}_{\rm tot} - \frac{i}{2} C_j L_j \right] \right) = \int D x_{Rj}' D x_{Lj}' \exp \left( - \sqrt{\lambda} T \T S_{R+L+C} \right) \, , \label{eq:Stilde-tot-def}
\end{equation}
where $\tilde{S}_{\rm tot} = \tilde{S}_{\rm tot}({\bf C})$ is a function of ${\bf C}$. With this definition, we now have
\begin{equation}
    \langle W[\mathcal{C}] \rangle_T = \int dC_j \exp \left(-\sqrt{\lambda} T \T \left[ \tilde{S}_{\rm tot} - \frac{i}{2} C_j L_j \right] \right) \, . \label{eq:W-loop-as-FT}
\end{equation}
Comparing this expression to Eq.~\eqref{eq:P-W-intro} makes it manifest that one should identify
\begin{equation}
    {\bf k} = \frac{\sqrt{\lambda} T \T }{2} {\bf C}
\end{equation}
as the momentum variable appearing in the momentum transfer distribution $P({\bf k})$. 
Then, we have that
\begin{equation}
    P({\bf k}) \propto \exp \left[ - \sqrt{\lambda} T \T \tilde{S}_{\rm tot}\left(  \frac{2 {\bf k} }{\sqrt{\lambda} T \T} \right) \right] \, . \label{eq:prob-dist}
\end{equation}
As such, our main object of interest in the subsequent Sections will be the evaluation of  $\tilde{S}_{\rm tot}({\bf C})=\tilde{S}_{\rm tot}(2{\bf k}/(\sqrt{\lambda}T\T))$.

At this point, it is worth discussing how the Wilson loop and $\tilde{S}_{\rm tot}$ are related in the strong coupling $\lambda \to \infty$ limit (in practice we only need $\sqrt{\lambda} T \T \to \infty$). Given this limit, one will have
\begin{equation}
    \langle W[\mathcal{C}] \rangle_T = \exp \left( - \sqrt{\lambda} T \T S_{\rm tot}({\bf L}) \right) \, , \label{eq:W-loop-Stot-L}
\end{equation}
where $S_{\rm tot}({\bf L})$ is a function of ${\bf L}$ obtained by extremizing $S_{\rm tot}$ defined in Eq.~\eqref{eq:S-tot} and evaluating it on the extremal surface with boundary conditions for the worldsheet specified by ${\bf L}$.  With the definition of $S_{\rm tot}({\bf L})$ in hand, we can now write Eq.~\eqref{eq:P-W-intro} from the Introduction as
\begin{equation}
    P({\bf k}) = \frac{1}{(2\pi)^3} \int d^3 {\bf L} \, e^{ - i {\bf k} \cdot {\bf L} } \exp \left( - \sqrt{\lambda} T \T S_{\rm tot}({\bf L}) \right) \, , \label{eq:P-from-S-first}
\end{equation}
which is the inverse Fourier transform of Eq.~\eqref{eq:W-loop-as-FT}.
We turn now to deriving the relationship between ${S}_{\rm tot}({\bf L})$ and $\tilde{S}_{\rm tot}({\bf C})$ starting from Eq.~\eqref{eq:W-loop-as-FT}, as we will find it useful later in Sections~\ref{sec:non-gaussian} and~\ref{sec:consequences}.

Carrying out the integral on the right hand side of  Eq.~\eqref{eq:W-loop-as-FT} over $C_i$ in the saddle point approximation, one obtains that the result is given by evaluating $\tilde{S}_{\rm tot}({\bf C}) + i C_j L_j/2$ at the value of $C_j$ determined by
\begin{equation}
    \frac{\partial}{\partial C_j} \left[ \tilde{S}_{\rm tot}({\bf C}) - \frac{i}{2} {\bf C} \cdot {\bf L} \right] = 0 \, .
\end{equation}
We denote the solution to this equation by $\bar{C}_j({\bf L})$. This can be recast as an implicit relation
\begin{equation}
    L_j = - 2i \left. \frac{\partial \tilde{S}_{\rm tot} }{\partial C_j} \right|_{{\bf C} = \bar{\bf C}(\bf L)} \, . \label{eq:L-C-relation}
\end{equation}
It then follows that 
\begin{align}
    S_{\rm tot}({\bf L}) = \left[ \tilde{S}_{\rm tot}({\bf C}) - \frac{i}{2} {\bf C} \cdot {\bf L} \right]_{{\bf C} = {\bf \bar{C}}({\bf L}) } = \left[ \tilde{S}_{\rm tot}({\bf C}) - {\bf C} \cdot \frac{\partial \tilde{S}_{\rm tot} }{\partial {\bf C} } \right]_{{\bf C} = {\bf \bar{C}}({\bf L}) } \, . \label{eq:S-Stot-relation}
\end{align}
In other words, there is a concrete sense in which $\tilde{S}_{\rm tot}({\bf C})$ and $S_{\rm tot}({\bf L})$ are Legendre transforms of each other. Given that $S_{\rm tot}({\bf L})$ is the one directly related to the action in a path integral, it is natural to think of $\tilde{S}_{\rm tot}({\bf C})$ as a Hamiltonian. This is consistent with the fact that, as required in order to have a probability interpretation of $P({\bf k})$, and as we will see by explicit calculations in what follows, $\tilde{S}_{\rm tot}({\bf C})$ is real and bounded from below. On the other hand, this is not necessarily the case for $S_{\rm tot}({\bf L})$, as it should already be clear from Eq.~\eqref{eq:L-C-relation} that $\bar{\bf C}({\bf L})$ will in general take complex values. It follows from this discussion that the worldsheet configurations that extremize $S_{R+L+C}$ in Eq.~\eqref{eq:W-loop-Stot} will in general be complex-valued. Consequently, except for the ${\bf L} = 0$ case, in which our calculation will reduce to that of Refs.~\cite{Gubser:2006bz,Herzog:2006gh}, it is not possible to associate a real classical  string worldsheet configuration with the Wilson loop we consider in this work for a general spatial separation ${\bf L}$ between the Wilson lines. 
That said, it is possible to find the associated complex string worldsheet -- as we have ourselves done explicitly before we realized that we do not need explicit expressions for the complex string worldsheet associated with $S_{\rm tot}({\bf L})$ in order to determine $\tilde S_{\rm tot}({\bf C})$ and the momentum transfer distribution $P({\bf k})$ explicitly$\ldots$

\section{The Momentum Transfer 
Distribution} \label{sec:momentum-broadening}

Based on the setup discussed in Section~\ref{sec:ads-cft}, in Section~\ref{sec:calculation} we derive an explicit expression for the momentum transfer distribution $P({\bf k})$ that lends itself to numerical evaluation. We then explain how known results for the heavy quark drag coefficient $\eta_D$, the heavy quark diffusion coefficients $\kappa_L$ and $\kappa_T$ and the jet quenching parameter $\hat{q}$ can all be obtained from  suitable Gaussian approximations to $P({\bf k})$. In Section~\ref{sec:non-gaussian}, we go beyond the known Gaussian characteristics of $P({\bf k})$. We characterize the previously undescribed non-Gaussian features of $P({\bf k})$ and we point to prominent, qualitatively novel, correlations between the longitudinal and transverse momentum fluctuations. 
These results may be used to formulate a complete first-principles description of 
heavy quark transport in
strongly coupled ${\cal N}=4$ plasma
including
the momentum transfer, stopping, diffusion and equilibration processes for a heavy quark that may initially be ultrarelativistic. We shall describe and discuss this in Section~\ref{sec:consequences}.

\subsection{Calculation and Results} \label{sec:calculation}

\subsubsection{The Bulk 
Contribution to \texorpdfstring{$\tilde{S}_{\rm tot}({\bf C})$}{Stot(C)}}

With Eqs.~\eqref{eq:SXR},~\eqref{eq:SXL}, and~\eqref{eq:S-tot} in hand, we carry out the integral in Eq.~\eqref{eq:W-loop-Stot} over $x_{Rj}'$ and $x_{Lj}'$ in the saddle point approximation by demanding $\delta S_{\rm tot}/\delta x_{Ri}'=0$ and $\delta S_{\rm tot}/\delta x_{Li}' = 0$. For simplicity, throughout the calculation we set units such that $\pi T = 1$. This yields the equations of motion
\begin{align}
    x_{R1}' &= i\frac{C_{1} z^2 }{f_- - v^2} \sqrt{ - \frac{f_-}{f} \frac{(f_- - v^2)^2}{(f_- - v^2) (f_- - C_{3}^2 z^4) - C_\perp^2 z^4 f_- }} \, , \label{eq:xR1prime} \\
    x_{L1}' &= -i\frac{C_{1} z^2 }{f_+ - v^2} \sqrt{ - \frac{f_+}{f} \frac{(f_+ - v^2)^2}{(f_+ - v^2) (f_+ - C_{3}^2 z^4) - C_\perp^2 z^4 f_+ }} \, , \label{eq:xL1prime} \\
    x_{R2}' &= i\frac{C_{2} z^2 }{f_- - v^2} \sqrt{ - \frac{f_-}{f} \frac{(f_- - v^2)^2}{(f_- - v^2) (f_- - C_{3}^2 z^4) - C_\perp^2 z^4 f_- }} \, , \label{eq:xR2prime} \\
    x_{L2}' &= -i\frac{C_{2} z^2 }{f_+ - v^2} \sqrt{ - \frac{f_+}{f} \frac{(f_+ - v^2)^2}{(f_+ - v^2) (f_+ - C_{3}^2 z^4) - C_\perp^2 z^4 f_+ }} \, , \label{eq:xL2prime} \\
    x_{R3}' &= i\frac{C_{3} z^2 }{f_-} \sqrt{ - \frac{f_-}{f} \frac{(f_- - v^2)^2}{(f_- - v^2) (f_- - C_{3}^2 z^4) - C_\perp^2 z^4 f_- }} \, , \label{eq:xR3prime} \\
    x_{L3}' &= -i\frac{C_{3} z^2 }{f_+} \sqrt{ - \frac{f_+}{f} \frac{(f_+ - v^2)^2}{(f_+ - v^2) (f_+ - C_{3}^2 z^4) - C_\perp^2 z^4 f_+ }} \, , \label{eq:xL3prime} 
\end{align}
where we have defined $C_\perp^2 \equiv C_1^2 + C_2^2$ for brevity. These equations in general have complex-valued solutions, which, on mathematical grounds, should only be allowed if the path integral in Eq.~\eqref{eq:W-loop-Stot} can be deformed into the complex plane without crossing any singularities. This is precisely the role that the $i\epsilon$'s in $S[X_R]$ and $S[X_L]$ have: to select the correct analytic continuation of the action when the integration paths for $x_{Rj}'$ and $x_{Lj}'$ are deformed into the complex plane via the expressions for $f_+$ and $f_-$ in Eq.~\eqref{eq:fpm}.

Keeping track of all of the $i\epsilon$'s, the above equations~\eqref{eq:xR1prime} --~\eqref{eq:xL3prime} can be solved algebraically. Inserting these solutions into the action integrals~\eqref{eq:SXR},~\eqref{eq:SXL} allows us to rewrite the square roots in the integrands of $S\lbrack X_R\rbrack$, $S\lbrack X_L\rbrack$:
\begin{align}
    \sqrt{ - \left( f_- x_{R3}'^2 + (f_- - v^2 ) \left( x_{R1}'^2 + x_{R2}'^2 + \frac1f \right) \right) } &=  \sqrt{ - \frac{f_-}{f} \frac{(f_- - v^2)^2}{(f_- - v^2) (f_- - C_{3}^2 z^4) - C_\perp^2 z^4 f_- }} \, , \label{eq:action-arg-R} \\
    \sqrt{ - \left( f_+ x_{L3}'^2 + (f_+ - v^2 ) \left( x_{L1}'^2 + x_{L2}'^2 + \frac1f \right) \right) } &= \sqrt{ - \frac{f_+}{f} \frac{(f_+ - v^2)^2}{(f_+ - v^2) (f_+ - C_{3}^2 z^4) - C_\perp^2 z^4 f_+ }} \, . \label{eq:action-arg-L}
\end{align}
With $S_{\rm tot} $ in \eqref{eq:S-tot} fully specified, we then use \eqref{eq:Stilde-tot-def} to arrive at an explicit expression for $\tilde{S}_{\rm tot}({\bf C})$. Exploiting 
that $x_{Rj}'$ and $x_{Lj}'$ are complex conjugates of each other,  $\tilde{S}_{\rm tot}({\bf C})$ can be written in terms of the real part of an expression that depends only on $f_+$ (or only $f_-$) 
\begin{align}
    \tilde{S}_{\rm tot}({\bf C}) &=  \int_0^{1} \frac{dz}{z^2} {\rm Re} \left\{ \sqrt{ - \frac{f_+}{f} \frac{(f_+ - v^2)^2}{(f_+ - v^2) (f_+ - C_{3}^2 z^4) - C_\perp^2 z^4 f_+ }} \right\} \nonumber \\
    & \,\,\, -  C_3^2 \int_0^1 dz \, z^2  {\rm Re} \left\{ \frac{1}{f_+} \sqrt{ - \frac{f_+}{f} \frac{(f_+ - v^2)^2}{(f_+ - v^2) (f_+ - C_{3}^2 z^4) - C_\perp^2 z^4 f_+ }} \right\} \nonumber \\
    & \,\,\, -  (C_1^2 + C_2^2) \int_0^1 dz \, z^2  {\rm Re} \left\{ \frac{1}{f_+ - v^2} \sqrt{ - \frac{f_+}{f} \frac{(f_+ - v^2)^2}{(f_+ - v^2) (f_+ - C_{3}^2 z^4) - C_\perp^2 z^4 f_+ }} \right\} \, .
\end{align}
Further algebraic simplifications allow us to write
\begin{align}
    \tilde{S}_{\rm tot}({\bf C}) = \int_0^1 dz \, {\rm Re} \left\{ \frac{  1  }{z^2 f_+ } \sqrt{ - \frac{f_+}{f} \left[ (f_+ - v^2) (f_+ - z^4 C_3^2) - C_\perp^2 z^4 f_+ \right] } \right\} \, . \label{eq:Stot-intermediate-result}
\end{align}
Note that at $z = 0$, the argument of the square root in \eqref{eq:Stot-intermediate-result} is $-1 + v^2 + \mathcal{O}(\epsilon)$. 
Therefore, in a neighborhood of $z=0$, the integrand is imaginary and the contribution to $\tilde{S}_{\rm tot}({\bf C})$ vanishes. Similarly, the argument of the square root is real and negative (up to $\mathcal{O}(\epsilon)$) as $z \to 1^-$, and there is thus also no contribution coming from the square root to $\tilde{S}_{\rm tot}({\bf C}) $ from the 
$z$ integral in \eqref{eq:Stot-intermediate-result} near $z=1$. The real part of the square root only receives contributions from the points that are in between the solutions of
\begin{equation}
    (f - v^2) (f - z^4 C_3^2) - C_\perp^2 z^4 f = 0 \, ,
\end{equation}
for real $C_1, C_2, C_3$. The solutions to this equation are
\begin{equation}
    z_{\pm}^4 = \frac{1+C_\perp^2+ (1-v^2)(1+C_3^2) }{2(1+C_\perp^2+C_3^2)} \pm \frac{\sqrt{ C_\perp^4 + 2 C_\perp^2 (v^2 + C_3^2 (1 - v^2) ) + (v^2 - (1-v^2) C_3^2 )^2 } }{2(1+C_\perp^2+C_3^2)} \, , \label{eq:zpm}
\end{equation}
and satisfy $0 \leq z_- \leq z_+ \leq 1$. 
The integral in Eq.~\eqref{eq:Stot-intermediate-result} receives a ``bulk'' contribution coming from $z_- < z < z_+$ as well as a contribution coming from the horizon at $z=1$ that we shall discuss in the next Subsection.
The bulk contribution to
Eq.~\eqref{eq:Stot-intermediate-result} takes the form 
\begin{equation} 
    \tilde{S}_{\rm tot}({\bf C})\Big\vert_{\rm bulk} = \int_{z_-}^{z_+}   \frac{dz}{z^2 f} \sqrt{ C_\perp^2 z^4 f - (f - v^2) (f - z^4 C_3^2)  }
\label{eq:first-bulk-stot}
\end{equation}
where $z_\pm$ are given in Eq.~\eqref{eq:zpm}. Remarkably, this final result \eqref{eq:first-bulk-stot}
matches the definition~\cite{Gradshteyn} of an Appell hypergeometric series $F_1(a,b_1,b_2;c;x,y)$, allowing us to write the bulk contribution to $\tilde{S}_{\rm tot}({\bf C})$ analytically:
\begin{equation}
     \tilde{S}_{\rm tot}({\bf C})\Big\vert_{\rm bulk} 
     =\sqrt{1+{\bf C}^2} \frac{\pi}{32}  \frac{(z_+^4 - z_-^4 )^2 }{z_-^5 (1 - z_-^4) }  F_1 \! \left( \frac32 , \frac54 , 1 ; 3 ; -\frac{z_+^4 - z_-^4}{z_-^4} , \frac{z_+^4 - z_-^4}{1 - z_-^4} \right)\ .\label{eq:bulk-stot}
\end{equation}
We shall evaluate and interpret this expression below, after completing the derivation of 
$\tilde{S}_{\rm tot}({\bf C})$ in the next Subsection.

\subsubsection{The Contribution to \texorpdfstring{$\tilde{S}_{\rm tot}({\bf C})$}{Stot(C)} from the Horizon}

In~\eqref{eq:bulk-stot}, we have specified that the $\tilde{S}_{\rm tot}({\bf C})$ evaluated in the previous subsection is the bulk contribution. The reason for this specification is a technical subtlety that turns out to be of great physical importance and that we have neglected so far for the sake of a simplified presentation. The point is that the integrand of expression~\eqref{eq:Stot-intermediate-result} has a single pole at the horizon $z=1$ that is generated by $1/f$. This leads to the following ambiguity: 
if one views the 
expression~\eqref{eq:Stot-intermediate-result} as a contour integral from $z=0$ to $z=1$ on one side of the SK contour and then back to $z=0$ on the other side, one has to make a choice as to whether to encircle the pole at $z = 1$ or not. One can check by contour analysis that, if present, the contribution obtained from encircling the pole will be given by 
\begin{equation}
    \tilde{S}_{\rm tot}({\bf C})\Big\vert_{\rm horizon} = \frac{\pi}{2} | v C_3 | \, ,
    \label{eq:s-horizon}
\end{equation}
corresponding to the residue of the integrand at $z=1$. There are two different lines of argument that both indicate that this contribution must be included, one mathematical and one physical.

On mathematical grounds, we begin by observing that the bulk contribution $\tilde{S}_{\rm tot}({\bf C})\vert_{\rm bulk}$ given by \eqref{eq:bulk-stot} is not differentiable at $C_3=0$, and that the resulting discontinuity in the derivative of $\tilde{S}_{\rm tot}({\bf C})\vert_{\rm bulk}$ is exactly the same as the discontinuity in the derivative of~\eqref{eq:s-horizon}. Also, 
examining the integrand in~\eqref{eq:Stot-intermediate-result}, we see that we must choose whether the complex integration path does or does not encircle the pole at $z=1$. Ensuring that there is no discontinuity in the derivative of
\begin{equation}
    \tilde{S}_{\rm tot}({\bf C})
    =\tilde{S}_{\rm tot}({\bf C})\Big\vert_{\rm bulk}
    +\tilde{S}_{\rm tot}({\bf C})\Big\vert_{\rm horizon}
    \label{eq:stot-bulk-plus-horizon}
\end{equation}
requires making one choice for $vC_3>0$ and the other choice for $vC_3<0$.\footnote{Note that when $C_3 = 0$ or $v=0$ there is no pole in the integrand at $z=1$ because the square root vanishes at the horizon. This means that when we look at the integral in~\eqref{eq:Stot-intermediate-result} as an analytic function of $C_3$, it is possible for the complex contour defined by $z$ to ``move over'' the pole smoothly precisely as $C_3$ crosses zero, and thus transition from a path that doesn't encircle $z=1$ to one that does.} This means that the contribution \eqref{eq:s-horizon} from the horizon 
should be multiplied by a Heaviside step function that is unity for one sign of the product $v C_3$ and zero for the other. In this work, we choose our conventions such that the maximum of the probability distribution is at $C_3 > 0$ (corresponding to the heavy quark losing longitudinal momentum to the plasma) when $v > 0$ (when the heavy quark has positive longitudinal momentum). With this 
choice, which foreshadows the physical argument below, we obtain
\begin{align}
    \tilde{S}_{\rm tot}({\bf C}) &= \sqrt{1+{\bf C}^2} \frac{\pi}{32}  \frac{(z_+^4 - z_-^4 )^2 }{z_-^5 (1 - z_-^4) }  F_1 \! \left( \frac32 , \frac54 , 1 ; 3 ; -\frac{z_+^4 - z_-^4}{z_-^4} , \frac{z_+^4 - z_-^4}{1 - z_-^4} \right) + \frac{\pi}{2} |v C_3| \theta( - v C_3 ) \, , \label{eq:stot-result}
\end{align}
where $z_\pm$ are given 
by~\eqref{eq:zpm}. We emphasize that upon handling the contribution to it coming from the horizon correctly, $\tilde{S}_{\rm tot}({\bf C})$ in Eq.~\eqref{eq:stot-result} is differentiable at $C_3=0$.

The same conclusion can be reached if one writes the differential equations that $\tilde{S}_{\rm tot}({\bf C})\vert_{\rm bulk}$ satisfies  (essentially, the differential equations that define $F_1$) as functions of ${\bf C}$, and then integrates these equations starting from a positive value of $C_3$. One finds that the ``horizon'' contribution, including the Heaviside step function, will be generated automatically as one integrates from $C_3 > 0$ to $C_3 < 0$. This leads to two solutions (related by $C_3 \to - C_3$), each continuously differentiable, and each with a single point where the probability distribution peaks. (This maximum is at a position determined by $C_3 = \pm \gamma v$, as we will see in the next Subsection.) These two solutions are not symmetric under an inversion of the sign of the longitudinal momentum, reflecting the fact that losing energy to the plasma is not equivalent to gaining energy from the plasma. The physical solution is then selected by demanding that when $k_L>0$ it should be more likely for the heavy quark to lose $k_L$ than to gain it; energy loss should be more likely than energy gain.

Interestingly, we see that the mathematically dictated addition of the contribution~\eqref{eq:s-horizon} resolves an objection that one could have raised about~\eqref{eq:bulk-stot} on physical grounds. Namely, the bulk contribution~\eqref{eq:bulk-stot} is symmetric under $C_3 \to - C_3$ while one expects on physical grounds that the momentum transfer distribution $P({\bf k}) $ in~\eqref{eq:prob-dist} and thus $\tilde{S}_{\rm tot}({\bf C}) $ should be asymmetric in $k_L = \tfrac{\sqrt{\lambda} T \T }{2} C_3$ since a heavy quark moving at velocity $v$ should be more likely to decelerate as it propagates through strongly coupled plasma than to accelerate along its direction of motion. It is solely the contribution from the horizon that introduces this asymmetry. Therefore, the mathematical choice of whether to encircle the pole at $z = 1$ may be viewed as the additional boundary condition needed to break time reversal invariance in the calculation in the way that is required in order to describe the physics of heavy quark transport in plasma. 

\subsubsection{Qualitative Features of the Momentum Transfer 
Distribution \texorpdfstring{$P({\bf k})$}{P(k)}}
\label{sec:qualitative-features}

\begin{figure}
    \centering
    \includegraphics[height=0.61\textwidth]{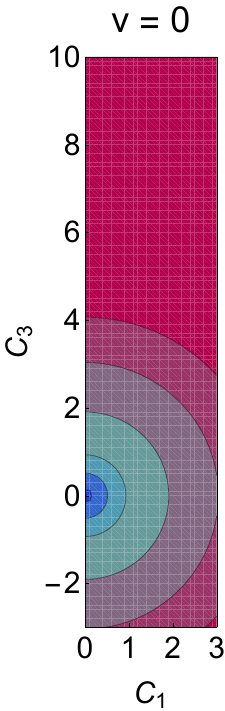}
    \includegraphics[height=0.61\textwidth]{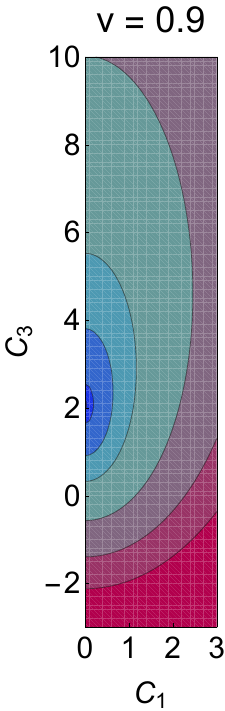}
    \includegraphics[height=0.61\textwidth]{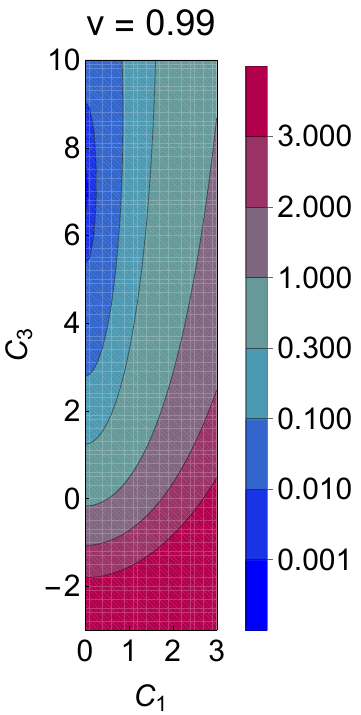}
    \includegraphics[height=0.61\textwidth]{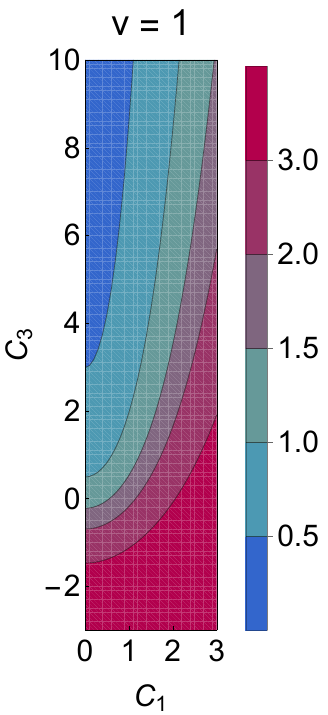}
    \caption{The logarithm of the momentum change distribution normalized by $- \sqrt{\lambda} \T T$, i.e., $\tilde{S}_{\rm tot}$, as a function of $C_1$ and $C_3$ for four different velocities $v=0$, 0.9, 0.99 and 1. We have set $C_2 = 0$, meaning that $C_1 = C_\perp$. We have used a common color scheme for $v \in \{0,0.9,0.99\}$, with logarithmically spaced contours at small values of $\tilde{S}_{\rm tot}$ so as to be able to visually identify the point of maximum probability, and linearly spaced contours at larger values. We have used a separate (linear) scale for $v=1$ as no point of maximum probability with finite $C_3$ exists in this case. 
    }
    \label{fig:stot}
\end{figure}

With equation \eqref{eq:stot-result} now in hand, we have an explicit analytical expression that determines the momentum transfer probability distribution to leading order in $\sqrt{\lambda}$:
\begin{equation}
    P({\bf k}) \propto \exp \left[ - \sqrt{\lambda} T \T \tilde{S}_{\rm tot}\left(  \frac{2 {\bf k} }{\sqrt{\lambda} T \T} \right) \right] \, . \label{eq:Prob-result}
\end{equation}
The normalization prefactor may be obtained by integrating over ${\bf k}$. Eqs.~\eqref{eq:stot-result} and \eqref{eq:Prob-result} are the main results of Section~\ref{sec:momentum-broadening}. The remainder of this paper constitutes an exploration of their features and consequences. 
And, in fact,
the resulting momentum transfer distribution for the case where the heavy quark has velocity $v=0.9$ was plotted already in Fig.~\ref{fig:P-of-k-intro}. In Fig.~\ref{fig:stot}, we plot the results of our calculation of 
$-\ln P({\bf k})/(\sqrt{\lambda}T{\cal T})$, 
corresponding to $\tilde{S}_{\rm tot}({\bf C})$, for four different values of the heavy quark velocity $v$.\footnote{To illustrate the discussion from the previous Subsection, we note that for $C_3>0$ the quantity $\tilde{S}_{\rm tot}({\bf C})$ is equal to 
$\tilde{S}_{\rm tot}({\bf C})\big\vert_{\rm bulk}$ given by the expression \eqref{eq:bulk-stot}, but if we had plotted the expression \eqref{eq:bulk-stot} in Fig.~\ref{fig:stot} the plots would have been symmetric about $C_3=0$ meaning that for $v>0$ they would have had a kink at $C_3=0$. Adding the horizon contribution \eqref{eq:s-horizon} modifies $\tilde{S}_{\rm tot}({\bf C})$ for $C_3<0$ when $v>0$ in such a way as to yield the (physical) results shown in the Fig.~\ref{fig:stot}.}

Several important qualitative  features of the momentum transfer probability distribution $P({\bf k})$ for a heavy quark propagating through strongly coupled plasma are now apparent from Fig.~\ref{fig:stot}, features which one can verify and quantify by explicit evaluation of the expression \eqref{eq:stot-result}. We list the most prominent ones below. At this point we restore units, inserting factors of $\pi T$ 
wherever appropriate.

\begin{enumerate}
    \item {\it The drag coefficient.}\\
    The minimum of $\tilde{S}_{\rm tot}$ is located at $C_1 = C_2 = 0$ and  $C_3 = \gamma v$  where $\gamma = (1 - v^2)^{-1/2}$ is the Lorentz boost factor. This determines the maximum of $P({\bf k})$, which is also the mean of the distribution in the large $\lambda$ limit, and which is located at 
    \begin{align}
        \langle k_1 \rangle = \langle k_2 \rangle = 0 \, , & & \langle k_3 \rangle = \frac{\pi \sqrt{\lambda} T^2 \T}{2} \gamma v \, 
    \end{align}
    The string configuration determined by these values of $C_i$ is exactly the trailing string of Refs.~\cite{Gubser:2006bz,Herzog:2006gh}, and it implies the drag coefficient $\eta_D=\langle k_3 \rangle /({\cal T} \gamma v M)$ identified therein.
    
    \item  {\it The heavy quark momentum diffusion coefficients $\kappa_T$ and $\kappa_L$.}\\
    The widths along $k_i$ of the momentum transfer distribution around its maximum or, equivalently, the variances of the distribution $P({\bf k})$ in the large $\lambda$ limit, are given by
    \begin{align}
        \langle k_1^2 \rangle = \langle k_2^2 \rangle &= \pi \T \sqrt{\lambda} \gamma^{1/2} T^3 \equiv \T \kappa_T \, , \\
        \langle (k_3 - \langle k_3 \rangle)^2 \rangle &= \pi \T \sqrt{\lambda} \gamma^{5/2} T^3 \equiv \T \kappa_L \, , 
    \end{align}
    which are exactly the results obtained first in Refs.~\cite{Gubser:2006nz,Casalderrey-Solana:2007ahi}. This follows from expanding Eq.~\eqref{eq:stot-result} up to quadratic order around the maximum of $P({\bf k})$:    
    \begin{equation}
        (\pi T)^2 \tilde{S}_{\rm tot}(C_\perp, C_3 = \gamma v + \delta C_3) = \frac{\pi}{2} \frac{ C_\perp^2}{4 \gamma^{1/2} } + \frac{\pi}{2} \frac{ (\delta C_3)^2}{4 \gamma^{5/2} } + \mathcal{O}(C_\perp,\delta C_3)^3 \, .
    \end{equation}
    \item {\it Non-Gaussian features and Transverse-Longitudinal-correlations.}\\
    Although taking the strict strong coupling limit $\lambda \to \infty$ at fixed $v$ makes the distribution more and more Gaussian, the distribution becomes increasingly non-Gaussian as $v$ is increased at any fixed large value of $\lambda$. This means that 
    corrections to a description of the motion of the heavy quark based on a drag force and transverse and longitudinal momentum diffusion coefficients must become more and more important as $v$ is increased. As we have a complete expression for $P({\bf k})$ and not just the location of its maximum (which specifies $\eta_D$) and the Gaussian variances about its maximum ($\kappa_T$ and $\kappa_L$), we can for the first time compute these corrections. These corrections include in particular qualitatively novel correlations between transverse and longitudinal momentum fluctuations that  cannot be encoded in any Gaussian approximation to the  fluctuations.  A complete presentation of these non-Gaussian features will be our focus in Subsection~\ref{sec:non-gaussian}.
    
    \item {\it The jet quenching parameter $\hat{q}$}\\
    The jet quenching parameter $\hat{q}$ was introduced first in perturbative QCD calculations of parton energy loss where a light parton close to the eikonal limit $v\to 1$ loses energy via medium-induced radiation of a gluon. Although this radiation was treated at weak coupling, it was immediately understood that the incident and outgoing parton and the radiated gluon all interact strongly with and exchange transverse momentum with the QGP medium, meaning that completing this weakly coupled calculation requires non-perturbative information about transverse momentum broadening in a situation in which a hard parton does not exchange longitudinal momentum with the medium. The jet quenching parameter $\hat q \equiv \langle k_\perp^2 \rangle/\T$ is the mean transverse momentum squared exchanged with the medium per distance travelled in this situation in which the hard parton does not lose energy. It arises in a weakly coupled QCD calculation, but can be defined at strong coupling for any gauge theory plasma. The jet quenching parameter for the strongly coupled $\mathcal{N}=4$ SYM plasma, which was first calculated in Refs.~\cite{Liu:2006he,Liu:2006ug,DEramo:2010wup}, is determined by the variance of the momentum transfer distribution $P({\bf k})$ along the $C_1$ direction on the $C_3 = 0$ line when $v = 1$. 
    To compute this quantity, one can start from the complete result in Eq.~\eqref{eq:stot-result}, set $v=1$ and show by explicit calculation that
    \begin{equation}
        \tilde{S}_{\rm tot} = \frac{\sqrt{\pi} \, \Gamma(5/4) }{2 \, \Gamma(7/4) } \frac{(1 + C_\perp^2)^{3/4}}{(1+ C_\perp^2 + C_3^2)^{1/4}}  {}_2 F_1 \left( 1, \frac14 ; \frac74 ;  \frac{1 + C_\perp^2}{1+ C_\perp^2 + C_3^2} \right) + \frac{\pi}{2} |C_3| \theta( - C_3 ) \, , \label{eq:Stot-lightlike}
    \end{equation}
    where ${}_2 F_1(a,b;c;x)$ is the ordinary hypergeometric function.\footnote{Note that (as discussed in Refs.~\cite{Gubser:2006nz,Casalderrey-Solana:2007ahi,Liu:2006he} and as we shall discuss in Section~\ref{sec:regime}) the calculation of $P({\bf k})$ breaks down if one attempts to take $v\to 1$ at fixed $M/T$ or if one takes $M/T\to 0$. Hence, evaluating $P({\bf k})$ with $v$ set to $1$ as we are doing here does not describe the transfer of momentum between a massless parton and the strongly coupled plasma of ${\cal N}=4$ SYM theory. The motivation to study $P({\bf k)}$ with $v=1$ at zero energy loss and in so doing to define $\hat{q}$ comes from weakly coupled QCD.  
    However, calculations of the mean energy loss rate~\cite{Chesler:2014jva,Chesler:2015nqz} for a massless particle in the strongly coupled $\mathcal{N}=4$ SYM plasma have been performed. They yield results in terms of the fraction of energy lost, which is a regime that is inaccessible in our present setup due to the kinematic limitation $|k_3| \ll E$. This  is not inconsistent with the fact that the maximum of the probability distribution determined by Eq.~\eqref{eq:Stot-lightlike} is at $C_3 \to \infty$ when $v=1$ and the fact that the distribution we calculated is not normalizable in this case.} 
    Restricting to $C_3=0$  and restoring units, one obtains
    \begin{equation}
        \tilde{S}_{\rm tot}(C_\perp, C_3 = 0) = \frac{\sqrt{\pi} \, \Gamma(5/4)}{\Gamma(3/4)} \sqrt{1 + \frac{C_\perp^2}{(\pi T)^2} }  \, , 
    \end{equation}
    and from here it is straightforward to show that
    \begin{equation}
        \langle k_1^2 \rangle_{k_3 = 0} = \langle k_2^2 \rangle_{k_3 = 0} = \frac{\pi \T \sqrt{\lambda} T^3 }{4} \frac{\sqrt{\pi} \, \Gamma(3/4) }{\Gamma(5/4)} \equiv \frac{1}{4} \T \hat{q} \, , \label{eq:qhat-result}
    \end{equation}
    which reproduces the result for $\hat q$ in strongly coupled ${\cal N}=4$ SYM theory first obtained in Refs.~\cite{Liu:2006he,Liu:2006ug,DEramo:2010wup}. The factor of $\frac{1}{4}$ in \eqref{eq:qhat-result} is the product of two factors of $\frac{1}{2}$. The first arises because
    $\langle k_\perp^2 \rangle = 2 \langle k_1^2 \rangle = 2\langle k_2^2\rangle$. The second arises because  $\hat{q}$, as calculated in Refs.~\cite{Liu:2006he,Liu:2006ug}, is defined in terms of $\langle k_\perp^2 \rangle$ for a (massless) parton in the adjoint representation whereas we have formulated our calculation for a (heavy) parton in the fundamental representation, with the factor of $\frac{1}{2}$ arising as the ratio between the result obtained in the fundamental and adjoint representations in the large-$N_c$ limit. Elsewhere in this paper, we focus on computing the (unrestricted) moments of $P({\bf k})$ and understanding their consequences for heavy quark transport; the extraction of $\hat q$ from $P({\bf k_\perp},k_3=0)$ at $v=1$ illustrates the fact that there is more physics to be found in the full $P({\bf k})$ than just what can be extracted from its moments.
\end{enumerate}

In summary, we have derived a distribution $P({\bf k})$ that provides a unified description of all the previously known Gaussian characteristics of the momentum transfer between a heavy quark and the strongly coupled plasma of ${\cal N}=4$ SYM theory 
(the heavy quark drag coefficient $\eta_D$, the heavy quark momentum diffusion coefficients $\kappa_L$, $\kappa_T$) as well as the jet quenching parameter $\hat{q}$. And, we now have at our finger tips a complete analytic expression for the momentum transfer distribution $P({\bf k})$ including complete information about all of its non-Gaussian features -- to which we now turn.

\subsection{Non-Gaussian Features and their Velocity Scalings} \label{sec:non-gaussian}

A straightforward way to characterize how the momentum transfer distribution $P({\bf k})$ deviates from Gaussianity is to determine its higher  order moments. The moments are defined as
\begin{equation}
    \langle k_1^{n_1} k_2^{n_2} k_3^{n_3} \rangle = \int d^3{\bf k} \, k_1^{n_1} k_2^{n_2} k_3^{n_3} P({\bf k}) \, ,
\end{equation}
which, with the help of Eq.~\eqref{eq:W-loop-as-FT}, can be written as
\begin{equation}
    \langle k_1^{n_1} k_2^{n_2} k_3^{n_3} \rangle = - (- i)^{n_1 + n_2 + n_3} \left. \frac{\partial^{n_1}}{\partial L_1^{n_1} } \frac{\partial^{n_2}}{\partial L_2^{n_2} } \frac{\partial^{n_3}}{\partial L_3^{n_3} } \langle W[\mathcal{C}] \rangle_T({\bf L}) \right|_{{\bf L} = 0} \, .
\end{equation}
As such, the \textit{connected}
moments of $P({\bf k})$ can be calculated by using the logarithm of the Wilson loop~\eqref{eq:W-loop-Stot-L} as their generating functional:
\begin{equation}
    \langle k_1^{n_1} k_2^{n_2} k_3^{n_3} \rangle_c = - (- i)^{n_1 + n_2 + n_3} \sqrt{\lambda} T \T \left. \frac{\partial^{n_1}}{\partial L_1^{n_1} } \frac{\partial^{n_2}}{\partial L_2^{n_2} } \frac{\partial^{n_3}}{\partial L_3^{n_3} } S_{\rm tot}({\bf L}) \right|_{{\bf L} = 0} \, . \label{eq:moments-from-S}
\end{equation}
We emphasize that it is the connected moments  that encode purely non-Gaussian information. Indeed,
starting at third order ($\sum_{i=1}^3 n_i \geq 3$), the higher connected moments would vanish if the distribution were Gaussian. However, as seen already qualitatively from Figure~\ref{fig:stot}, the higher moments do not vanish. 
These moments contain a wealth of information, as they reveal, for instance, how heavy quark energy loss is correlated with momentum broadening, or how transverse momentum broadening is correlated with longitudinal momentum broadening, or the skewness and kurtosis and all higher moments of the fluctuations of the transverse and longitudinal momenta. 
Here we discuss the resulting physics in more detail.  

\subsubsection{Moments of the Momentum Transfer Distribution} \label{sec:moments}

Equation~\eqref{eq:moments-from-S} determines the connected moments in terms of derivatives of $S_{\rm tot}({\bf L})$ which, following Eq.~\eqref{eq:S-Stot-relation}, can be expressed in terms of $\tilde{S}_{\rm tot}({\bf C})$. This relation is valid to leading order in the strong coupling expansion.\footnote{ The curious reader may wonder whether it could be easier to determine the moments by integrating directly over $P({\bf k})$. This is possible, of course, but doing so would yield a result that also contains some, but not all, subleading contributions in $1/(\sqrt{\lambda} \T)$.  
A complete determinination of the moments to subleading order $1/(\sqrt{\lambda} \T)$
would require including corrections to the saddle point approximation that we have not computed. The method that we follow in this Section has the advantage that it automatically selects the leading contributions $1/(\sqrt{\lambda} \T)$ without including any subleading terms, and allows us to calculate the moments analytically in a complete manner up to this order in the strong coupling expansion.} Given that we need to take derivatives of $S_{\rm tot}({\bf L})$, and since our explicit results are functions of ${\bf C}$, we need to evaluate $\partial C_k/\partial L_j$. With the help of Eq.~\eqref{eq:L-C-relation} we obtain 
\begin{equation}
    \frac{\partial C_k}{\partial L_j} = - \frac{1}{2i} \left[ \frac{\partial^2 \tilde{S}_{\rm tot} }{\partial C \partial C} \right]_{jk}^{-1} \, ,
\end{equation}
where $[\partial^2 \tilde{S}_{\rm tot}/\partial C \partial C]_{jk}^{-1}$ is the $jk$-th matrix element in the inverse of the Hessian matrix $\partial^2 \tilde{S}_{\rm tot}({\bf C})/\partial C_j \partial C_k$. We can thus write the connected moments of the distribution as
\begin{align}
    \langle k_1^{n_1} k_2^{n_2} k_3^{n_3} \rangle_c = -  \sqrt{\lambda} T \T & \left( \frac12 \left[ \frac{\partial^2 \tilde{S}_{\rm tot} }{\partial C \partial C} \right]_{1k}^{-1} \frac{\partial}{\partial C_k } \right)^{n_1} \!\!\! \times \left( \frac12 \left[ \frac{\partial^2 \tilde{S}_{\rm tot} }{\partial C \partial C} \right]_{2k}^{-1} \frac{\partial}{\partial C_k } \right)^{n_2} \nonumber \\   \times & \left. \left( \frac12 \left[ \frac{\partial^2 \tilde{S}_{\rm tot} }{\partial C \partial C} \right]_{3k}^{-1} \frac{\partial}{\partial C_k } \right)^{n_3} \!\!\! \times \left[ \tilde{S}_{\rm tot}({\bf C}) - {\bf C} \cdot \frac{\partial \tilde{S}_{\rm tot} }{\partial {\bf C} } \right] \right|_{C_1 = C_2 = 0 \, , C_3 = \pi T \gamma v} \, . \label{eq:moments-from-tStot}
\end{align}
Given that we have an explicit expression for $\tilde{S}_{\rm tot}$, this is a lengthy but conceptually straightforward expression to evaluate, evading the need to calculate $S_{\rm tot}({\bf L})$ altogether. In practice, the most efficient way of calculating the moments using Eq.~\eqref{eq:moments-from-tStot} is to expand $\tilde{S}_{\rm tot}$ up to sufficiently high order around $C_1 = C_2 = 0$, $C_3 = \pi T \gamma v$ (the order of the expansion needs to be higher than the number of derivatives to be taken) and then simply take the derivatives. 

By evaluating the expression \eqref{eq:moments-from-tStot} for $(n_1,n_2,n_3)=(0,0,1)$, $(2,0,0)$ and $(0,2,0)$, we have reproduced the classic results for the drag and the momentum diffusion coefficients discussed in the preceding section. The same equation~\eqref{eq:moments-from-tStot} allows us to derive the following novel results for the lowest non-Gaussian moments:
\begin{enumerate}
    \item The nonzero connected third-order moments of the momentum transfer distribution are given by:
    \begin{align}
        \langle (k_3 - \langle k_3 \rangle) k_1^2 \rangle = \langle (k_3 - \langle k_3 \rangle) k_2^2 \rangle &= \pi \sqrt{\lambda} \T \gamma^2 T^4 \frac{3v}{2} \, , \label{eq:Eloss-broadening-corr} \\  \langle (k_3 - \langle k_3 \rangle )^3 \rangle &= \pi \sqrt{\lambda} \T \gamma^4 T^4 \frac{9v}{2} \, . \label{eq:skewness}
    \end{align}
    \item The nonzero connected fourth-order moments of the momentum transfer distribution are given by:
    \begin{align}
        \langle k_1^4 \rangle - 3\langle k_1^2 \rangle^2 = \langle k_2^4 \rangle - 3\langle k_2^2 \rangle^2 &= \pi \sqrt{\lambda} \T \gamma^{3/2} T^5 \frac{9(8-v^2)}{8} \, , \label{eq:kurtosis-k1k2} \\
        \langle k_1^2 k_2^2 \rangle - \langle k_1^2 \rangle \langle k_2^2 \rangle &= \pi \sqrt{\lambda} \T \gamma^{3/2} T^5 \frac{3(8-v^2)}{8} \, , \label{eq:cross2-k1k2} \\
        \langle k_1^2 (k_3 - \langle k_3 \rangle)^2 \rangle  - \langle k_1^2 \rangle \langle (k_3 - \langle k_3 \rangle)^2 \rangle   &= \pi \sqrt{\lambda} \T \gamma^{7/2} T^5 \frac{3(8+9v^2)}{8} \, , 
        \label{eq:cross2-k1k3}\\
        \langle k_2^2 (k_3 - \langle k_3 \rangle)^2 \rangle  - \langle k_2^2 \rangle \langle (k_3 - \langle k_3 \rangle)^2 \rangle  &= \pi \sqrt{\lambda} \T \gamma^{7/2} T^5 \frac{3(8+9v^2)}{8} \, , 
        \label{eq:cross2-k2k3}\\
        \langle (k_3 - \langle k_3 \rangle)^4 \rangle - 3 \langle (k_3 - \langle k_3 \rangle)^2 \rangle^2 &= \pi \sqrt{\lambda} \T \gamma^{11/2} T^5 \frac{9(8+19v^2)}{8} \, .
        \label{eq:kurtosis-k3}
    \end{align}
\end{enumerate}
We can readily calculate 
higher-order moments by taking more derivatives in Eq.~\eqref{eq:moments-from-tStot}. 
One may also obtain the moments where both $k_1$ and $k_2$ appear simultaneously by introducing the azimuthal angle $\phi$, with $k_1 = k_\perp \cos \phi$ and $k_2 = k_\perp \sin \phi$, and using the fact that $P({\bf k})$ does not depend on $\phi$ to relate the moments where only $k_1$ appears to the moments where both $k_1$ and $k_2$ may appear. This is already apparent by comparing Eqs.~\eqref{eq:kurtosis-k1k2} and~\eqref{eq:cross2-k1k2}.

These results encode qualitative aspects of the physics of heavy quark propagation in QGP that have not received attention so far. For example, the expression \eqref{eq:Eloss-broadening-corr}) shows that
the energy loss distribution has positive skewness, meaning that the tail of the distribution at large positive values of $k_3$ falls off more slowly than at negative values of $k_3$ -- a feature that can be seen at a qualitative level by inspection of Fig.~\ref{fig:P-of-k-intro} or~\ref{fig:stot}. The bigger the skewness, the more asymmetric the energy loss probability is, meaning that as $\gamma$ increases $\kappa_L$ at some point stops being an adequate quantifier of longitudinal momentum broadening. 
As another example, we see from the expressions~\eqref{eq:kurtosis-k1k2} and~\eqref{eq:kurtosis-k3} that the kurtosis of both the transverse and longitudinal momentum fluctuations are positive, that both grow with increasing $\gamma$, but that the kurtosis of the longitudinal momentum fluctuations grows particularly rapidly. This again suggests that as $\gamma$ increases the longitudinal momentum transfer distribution cannot be characterized by $\kappa_L$ alone. 
Furthermore, we see from \eqref{eq:Eloss-broadening-corr} and \eqref{eq:cross2-k1k3} that energy loss and transverse momentum broadening are correlated, with more broadening being more likely if the energy loss is larger. This is qualitatively clear from Figure~\ref{fig:stot}, but it is now quantified by Eqs.~\eqref{eq:Eloss-broadening-corr} and \eqref{eq:cross2-k1k3}. The fact that these correlations grow faster with $\gamma$ than the products $\langle k_3 \rangle \langle k_\perp^2 \rangle$ 
and $\langle k_3\rangle^2 \langle k_\perp^2\rangle$, respectively, makes it extremely interesting to study their consequences for heavy quark transport.

\subsubsection{Scaling of the 
Moments and their Derivation in the Rest Frame of the Heavy Quark} \label{sec:HQframe}

It is natural to ask whether the higher moments calculated here display any discernible pattern from which we may draw intuition. In this context, we note that the moments explicitly written in Section~\ref{sec:moments} show the following scaling behavior in the limit $v\to 1$:
\begin{equation} \label{eq:scaling-pattern-QGPframe}
    \langle k_\perp^{2m} \, k_3^n \rangle_c \propto \sqrt{\lambda} \T (\sqrt{\gamma} \,  T)^{n + 2m + 1} \times \gamma^{n - 1} \, .
\end{equation}
Here the subscript $\langle \cdot \rangle_c$ means that the connected component of the moments is to be taken, and $k_\perp^{2m}$ may be any combination of even powers of $k_1, k_2$ adding up to $2m$. We now explain the reason for this scaling and show that it holds for any $m$ and $n$, not just for the explicit examples given in Section~\ref{sec:moments}.

Our derivation of Eq.~\eqref{eq:stot-result} and in fact all of our calculations so far were carried out in the rest frame of the plasma. However, the momentum transfer distribution depends on the coordinate frame in which we follow the momentum variables. As it turns out, the scaling pattern in Eq.~\eqref{eq:scaling-pattern-QGPframe} can be most naturally interpreted in the rest frame of the heavy quark. The connection between the velocity dependence of the moments in each frame will be fully fixed by the fact that the Wilson loop is a Lorentz scalar.  

To understand ~\eqref{eq:scaling-pattern-QGPframe}, we therefore go to the rest frame of the heavy quark whose coordinates we shall denote with primes:
\begin{align}
    t' = \gamma (t - v x_3 ) \, , & & x_3' = \gamma (x_3 - v t) \, . \nonumber
\end{align}
The two antiparallel lines that comprise the Wilson loop are located at $x_3 = vt$ and $x_3 = v t + L_3$, which uniquely fixes their $x_3'$ coordinates to be $0$ and $\gamma L_3$, meaning that the spatial separation in these new coordinates $L_3'$, conjugate to the momentum transfer in this frame, is given by $L_3' = \gamma L_3$. Also, the temporal extent of the loop along $t'$ is set by how much time passes on the line $x_3 = v t$, meaning $\T' = \gamma (\T - v^2 \T) = \T / \gamma$. The transverse separations $L_1, L_2$ do not change.
The momentum transfer distribution $P_{\rm HQ}({\bf C}')$ in the rest frame of the heavy quark can then be obtained by Lorentz transforming all the functional dependencies of Eqs.~\eqref{eq:stot-result} and~\eqref{eq:Prob-result} from the plasma rest frame to the heavy quark rest frame:
\begin{align}
    P_{\rm HQ}({\bf C}') &= \exp \left( - \sqrt{\lambda} T \gamma \T'  \tilde{S}_{\rm tot}( C_1 = C_1', C_2 = C_2', C_3 = \gamma C_3' ) \right) \, . \label{eq:HQ-frame-Prob}
\end{align}
In particular, we have used here   $\T = \gamma \T'$,  $C_3 = \gamma C_3'$ and thus 
\begin{equation} 
{\bf k}' = \frac{\sqrt{\lambda} T \gamma \T' }{2} {\bf C}' \, . \label{eq:HQ-frame-momentum-coords}
\end{equation} 
The functional form of $\tilde{S}_{\rm tot}$ does not change under this Lorentz transformation.

As we will demonstrate in a moment, the distribution in this frame exhibits a very particular feature in that all moments of $P_{\rm HQ}({\bf C}')$ depend on $\gamma$ and $T$ only via powers of $\sqrt{\gamma} \, T$ in the limit $v \to 1$. 
For the lowest order moments, this can be checked explicitly by expanding $\tilde{S}_{\rm tot}$ around $C_3' = v$, corresponding to 
\begin{align}
    \langle k_3' \rangle = \frac{\pi \sqrt{\lambda} T^2 \T' }{2} \gamma v \, .
\end{align}
One obtains
\begin{align}
    &- \sqrt{\lambda} T \gamma \T' \tilde{S}_{\rm tot}( C_1 = C_1', C_2 = C_2', C_3 = \gamma (v + \delta C_3') ) \nonumber \\
    &= - \frac{\sqrt{\lambda} \gamma t' }{\pi} \left[ \frac{\pi}{2} \frac{C_\perp'^2}{4 \gamma^{1/2}} + \frac{\pi}{2} \frac{(\delta C_3')^2}{4 \gamma^{1/2} } + \mathcal{O}(C_\perp',\delta C_3')^3 \right] \, ,
\end{align}
which implies that
\begin{align}
    \langle k_1'^2 \rangle = \langle k_2'^2 \rangle &= \pi \T' \sqrt{\lambda} \gamma^{3/2} T^3  \equiv {\cal T}'\kappa'_T
    \label{eq:kappaTprime}
    \, , \\
    \langle (k_3' - \langle k_3' \rangle)^2 \rangle &= \pi \T' \sqrt{\lambda} \gamma^{3/2} T^3 
    \equiv {\cal T}'\kappa'_L\, .
    \label{eq:kappaLprime}
\end{align} We see that in the rest frame of the heavy quark the variance of both the longitudinal and transverse 
momentum fluctuations
scale like $(\sqrt{\gamma} \, T)^3$ even though in the QGP rest frame $\kappa_T\propto \gamma^{1/2}$ and $\kappa_L \propto \gamma^{5/2}$.

To verify that all higher moments also scale with powers of $\sqrt{\gamma} \, T$, we rewrite  Eq.~\eqref{eq:moments-from-tStot} in terms of 
the variables $\delta C_1'$, $\delta C_2'$ and $\delta C_3'$, obtained from Lorentz transforming the $C_i$'s into the heavy quark rest frame, 
\begin{align}
    C_1 = C_1'= \delta C_1' \, , & & C_2 = C_2' = \delta C_2' \, , & & C_3 = \gamma C_3' = \gamma v + \gamma \delta C_3' \, .
\end{align}
To ease the notation, we again choose units such that $\pi T = 1$, and we define $ \tilde{S}_{\rm tot}'({\bf C}') = \sqrt{\gamma} \tilde{S}_{\rm tot}({\bf C}) $. 
We then find 
\begin{align}
    \langle k_1^{n_1} k_2^{n_2} k_3^{n_3} \rangle_c = -  \frac{\sqrt{\lambda} \T}{\pi \sqrt{\gamma}} & \left( \frac{\gamma^{1/2}}{2} \left[ \frac{\partial^2 \tilde{S}_{\rm tot}' }{\partial C' \partial C'} \right]_{1k}^{-1} \frac{\partial}{\partial C'_k } \right)^{n_1} \!\!\! \times \left( \frac{\gamma^{1/2}}{2} \left[ \frac{\partial^2 \tilde{S}_{\rm tot}' }{\partial C' \partial C'} \right]_{2k}^{-1} \frac{\partial}{\partial C'_k } \right)^{n_2} \nonumber \\   \times & \left. \left( \frac{\gamma^{3/2} }{2} \left[ \frac{\partial^2 \tilde{S}_{\rm tot}' }{\partial C' \partial C'} \right]_{3k}^{-1} \frac{\partial}{\partial C'_k } \right)^{n_3} \!\!\! \times \left[ \tilde{S}_{\rm tot}'({\bf C}') - {\bf C}' \cdot \frac{\partial \tilde{S}_{\rm tot}' }{\partial {\bf C}' } \right] \right|_{\delta {\bf C}' =0 } \nonumber \\
    = -  \frac{\sqrt{\lambda} \T}{\pi} \frac{\sqrt{\gamma}^{n_1+n_2+n_3+1} \gamma^{n_3-1}}{2^{n_1+n_2+n_3}}  & \left( \left[ \frac{\partial^2 \tilde{S}_{\rm tot}' }{\partial C' \partial C'} \right]_{1k}^{-1} \frac{\partial}{\partial C'_k } \right)^{n_1} \!\!\! \times \left(  \left[ \frac{\partial^2 \tilde{S}_{\rm tot}' }{\partial C' \partial C'} \right]_{2k}^{-1} \frac{\partial}{\partial C'_k } \right)^{n_2} \nonumber \\   \times & \left. \left(  \left[ \frac{\partial^2 \tilde{S}_{\rm tot}' }{\partial C' \partial C'} \right]_{3k}^{-1} \frac{\partial}{\partial C'_k } \right)^{n_3} \!\!\! \times \left[ \tilde{S}_{\rm tot}'({\bf C}') - {\bf C}' \cdot \frac{\partial \tilde{S}_{\rm tot}' }{\partial {\bf C}' } \right] \right|_{\delta {\bf C}' =0 }  \, .
    \label{eq:moments-from-tStot-resc}
\end{align}
Since the velocity scalings of Eq.~\eqref{eq:scaling-pattern-QGPframe} are given exactly by the prefactor $\sqrt{\gamma}^{n_1+n_2+n_3+1} \gamma^{n_3-1}$ (noting that in Eq.~\eqref{eq:scaling-pattern-QGPframe} we had set $n_3=n$ and had set $n_1$ or $n_2$ to $2m$), all that one needs to check is that $\tilde{S}_{\rm tot}'$ converges to a finite result when holding the value of $\delta {\bf C}'$ fixed. Indeed, one finds
\begin{align}
    & \lim_{v \to 1} \tilde{S}_{\rm tot}' = \nonumber \\
    &  \frac{ (\delta {\bf C}')^2 ( (\delta C_\perp')^2 + (\delta C_3' + 2)^2 ) \, {}_2 F_1 \! \left( \frac54 , \frac32 , 3 , - \frac{2 \sqrt{(\delta {\bf C})^2 ( (\delta C_\perp')^2 + (\delta C_3' + 2)^2 ) } }{1 + (\delta C_\perp')^2 + (1 + \delta C_3')^2 - \sqrt{(\delta {\bf C}')^2 ( (\delta C_\perp')^2 + (\delta C_3' + 2)^2 )}} \right) }{2^{15/4} \left( (1 + \delta C_3')^{2} \right)^{1/4} \left( 1 + (\delta C_\perp')^2 + (1 + \delta C_3')^2 - \sqrt{(\delta {\bf C}')^2 ( (\delta C_\perp')^2 + (\delta C_3' + 2)^2 )} \right)^{5/4} }
\end{align}
where $\delta C_\perp' = \sqrt{\delta C_1'^2 + \delta C_2'^2}$. As a consistency check, one may verify by taking derivatives that this expression reproduces the numerical values of the moments we calculated in the previous section in the limit $v \to 1$.
In summary, Eq.~\eqref{eq:moments-from-tStot-resc} establishes the velocity scaling in~\eqref{eq:scaling-pattern-QGPframe}  for all connected higher-order moments, with any values of $n$ and $m$. Restoring factors of $\pi T$ completes the derivation of Eq.~\eqref{eq:scaling-pattern-QGPframe}.

The velocity scaling of the connected moments~\eqref{eq:scaling-pattern-QGPframe} in the QGP rest frame is striking, but the pattern that this represents only becomes clear after we transform to the rest frame of the heavy quark, a frame in which it is the QGP that is moving with velocity $v$ and boost $\gamma$. 
In the heavy quark rest frame, one can see that by using the relation between the momenta in the two frames encoded in Eqs.~\eqref{eq:HQ-frame-Prob} and~\eqref{eq:HQ-frame-momentum-coords} and following the same reasoning that led us to Eqs.~\eqref{eq:kappaTprime} and~\eqref{eq:kappaLprime}, the scaling of the connected higher moments takes the form  
\begin{equation}
    \langle k_\perp^{2m} \, k_3^n \rangle_{c,{\rm HQ}} \propto \sqrt{\lambda} \T' (\sqrt{\gamma} \, T)^{n + 2m + 1} \, , \label{eq:scaling-pattern-HQframe}
\end{equation}
This makes it explicit that the momentum transfer distribution in the heavy quark rest frame depends on only one energy scale given by $\sqrt{\gamma} \, T$. This scale may be interpreted as an effective temperature that the heavy quark experiences as 
the strongly coupled QGP liquid flows past it at
ultrarelativistic speeds. This is precisely the velocity-dependent
effective temperature experienced by a heavy quark or a heavy quark-antiquark pair in a ``hot wind'' of flowing strongly coupled plasma found previously in Refs.~\cite{Liu:2006nn,Liu:2006he,Casalderrey-Solana:2007ahi,Gubser:2006nz,Chernicoff:2006hi,Ejaz:2007hg,Nijs:2023dbc}. Furthermore, it is remarkable to note that upon identifying the effective temperature as $\sqrt{\gamma} \, T$, the result of the fluctuation-dissipation theorem
\begin{equation}
    \kappa_L' = \kappa_T' \equiv \kappa' = 2 M (\sqrt{\gamma} \, T) \eta_D'
\end{equation}
holds, even though this system is very far from equilibrium. 
Note, however, that it holds only in this frame as a consequence of the identification of the effective temperature just mentioned, and \textit{not} in the QGP rest frame, which is where the stopping and equilibration of a heavy quark are naturally described.

\subsubsection{Quantifying the 
Relative Importance of Non-Gaussian Fluctuations and the Regime of Validity of their Calculation} \label{sec:regime}

Non-Gaussian features of $P({\bf k})$  become suppressed in the 
strict infinite-coupling limit, but they become enhanced for relativistic heavy quark velocities, making them potentially interesting from a phenomenological perspective. From a physics point of view, clarifying their relative importance is relevant since the Langevin picture for heavy quark transport (which was first discussed in the context of strongly coupled $\mathcal{N}=4$ SYM in Refs.~\cite{Gubser:2006nz,Casalderrey-Solana:2006fio,Casalderrey-Solana:2007ahi}) relies on the distribution $P({\bf k})$ being well-approximated 
as a Gaussian in transverse momentum and as a Gaussian in longitudinal momentum about the mean energy loss -- with no significant higher moments or correlations. 
However, even though doing so would appear to enhance the non-Gaussian features of $P({\bf k})$, we cannot take $\gamma$ to be arbitrarily large ($v$ arbitrarily close to 1) 
because our calculation breaks down at high velocity, as is also the case for previous treatments of heavy quarks or quark-antiquark pairs moving through strongly coupled plasma~\cite{Liu:2006nn,Liu:2006he,Casalderrey-Solana:2007ahi,Gubser:2006nz,Ejaz:2007hg,BitaghsirFadafan:2008adl}. In our case,
the calculation of the moments of the distribution breaks down at too high velocities because the magnitude of the typical momentum fluctuations that they describe fall outside the regime of validity of the effective field theory treatment upon which our calculation is built, as we now explain. Only after doing so can we assess the range of velocities where the Langevin picture is a good approximation and the range
of velocities where non-Gaussian features of $P({\bf k})$ are significant.

The regime of applicability of the EFT is restricted to total momentum transfer that is much smaller than the hard scale given by the mass of the heavy quark $M$. For transverse momentum broadening, the terms that are omitted in the HQET Lagrangian will scale as $k_\perp^2/M$ (resp. $k_\perp^2/Q$ for SCET) and the EFT derivation should thus apply for $\langle k_\perp^2 \rangle \T \sim \kappa_T \T^2 \ll M$. The corresponding condition for the longitudinal case is $\kappa_L \T^2 \ll \gamma^2 M$ (because the EFT constraint  $|k_3'| \ll M$ is formulated in the HQ rest frame, and $k_3'=k_3/\gamma$), which reduces to the same criterion as in the transverse case. In addition, the validity of our derivation relies upon taking $\T$ sufficiently large  that the calculation is dominated by the contribution to the Wilson loop that is extensive  in $\T$. This translates to $\T T \gg 1$, which is a stronger condition than $\sqrt{\lambda} T \T \gg 1$ (the necessary condition to take the saddle point approximation in our calculation). These parametric conditions combine to
\begin{equation}
    T^{-1} \ll \T \ll \sqrt{\frac{M}{\kappa_T}} \, , \label{eq:condition-Langevin}
\end{equation}
where the first inequality ensures that finite time effects have died away, and the second that the EFT description is applicable. 
Note that the hierarchy between the smallest and largest scale in~\eqref{eq:condition-Langevin} is equivalent to the ``speed limit''
\begin{equation}
    \sqrt{\gamma} \ll \frac{M}{\sqrt{\lambda} T} \, , \label{eq:speed-limit}
\end{equation}
which is the same inequality that was discussed in Refs.~\cite{Liu:2006nn,Liu:2006he,Casalderrey-Solana:2007ahi,Gubser:2006nz,Ejaz:2007hg,BitaghsirFadafan:2008adl} as a necessary condition for the calculations described therein to be valid. We see that, as we foreshadowed in Section~\ref{sec:qualitative-features}, our calculation of the moments would break down if we attempted either to take $v\to 1$ 
at fixed $M/T$ or to take $M/T\to 0$.

With an understanding of how the regime of validity of our calculation limits our ability to take $\gamma\to\infty$ in hand, we can 
now turn to assessing the regime of validity of the Langevin description.
The Langevin description is constructed by gluing together individual steps of time evolution satisfying the condition~\eqref{eq:condition-Langevin}, updating the new central value of the energy/longitudinal momentum at each step.
However, the Langevin description as discussed in~\cite{Gubser:2006nz,Casalderrey-Solana:2006fio,Casalderrey-Solana:2007ahi} only involves the drag and broadening coefficients, omitting all higher order moments of the momentum transfer distribution $P({\bf k})$. This amounts to assuming that $P({\bf k})$ is approximately Gaussian. Our calculation of the complete distribution $P({\bf k})$ now gives us the ability to ascertain where this Gaussian assumption breaks down.

The non-Gaussianity of the distribution $P({\bf k})$ may be quantified by taking ratios of the 
connected moments $\langle k_1^{2m} k_3^n \rangle_c$ to appropriate powers of the variances, concretely $\langle k_1^2 \rangle_c^m \langle k_3^2 \rangle_c^{n/2}$. We may evaluate this ratio using the parametric dependencies we have found in~\eqref{eq:scaling-pattern-QGPframe}:
\begin{equation}
    \frac{\langle k_1^{2m} k_3^n \rangle_c}{\langle k_1^2 \rangle_c^m \langle k_3^2 \rangle_c^{n/2}} \propto \frac{1}{(\sqrt{\lambda} T \T)^{m + n/2 - 1}} \frac{\gamma^{m+3n/2-1/2} }{ \gamma^{m/2 + 5n/4} } = \left(\frac{ \sqrt{\gamma} }{\sqrt{\lambda} T \T} \right)^{m + n/2 - 1} \equiv r^{m+n/2-1} \, , \label{eq:NG-criterion}
\end{equation}
where we have introduced $r = \sqrt{\gamma}/(\sqrt{\lambda} T \T)$ as the parameter that controls deviations from Gaussianity. 
Eq.~\eqref{eq:NG-criterion} quantifies what we have stated already above: the distribution becomes more and more non-Gaussian as the velocity of the heavy quark increases, but it becomes more and more Gaussian in the strong coupling and large time limit $\sqrt{\lambda} T \T \gg 1$. 
The momentum transfer distribution $P({\bf k})$ is well-approximated as Gaussian (and a Langevin approach is possible) as long as $r\ll 1$; non-Gaussian features of $P({\bf k})$ become significant at larger $r$, and are unsuppressed where $r\sim 1$ or greater.

As we have noted, however, the time $\T$ may not be taken to be arbitrarily large at each microscopic step in setting up our EFT calculation that yields the result~\eqref{eq:NG-criterion}. Turning the constraints \eqref{eq:condition-Langevin} on $\T$ into constraints on the dimensionless parameter $r$ that quantifies non-Gaussianity, we find that the regime of validity of our calculation corresponds to the regime
\begin{equation}
    \frac{\sqrt{\gamma}}{\sqrt{\lambda}} \left( \frac{M}{\sqrt{\gamma} \sqrt{\lambda} T} \right)^{-1/2}  < r < \frac{\sqrt{\gamma}}{\sqrt{\lambda}} \  . \label{eq:bound}
\end{equation}
The lower bound corresponds to the longest time $\T$ one could use from Eq.~\eqref{eq:condition-Langevin}. While the central limit theorem ensures that the distribution will approach a Gaussian at late times $\T$, one sees from the lower bound on $r$ in~\eqref{eq:bound} that 
even at the longest times ${\cal T}$
at which our calculation can be employed $r$ does not vanish, which means that  
non-Gaussian effects are present at some level. 
The upper bound on $r$ in~\eqref{eq:bound} 
is more interesting.  It tells us that if 
\begin{equation}
        \gamma > \lambda \,  \label{eq:NG-criterion-n}
\end{equation} 
then our calculation is valid in a regime in which $r>1$ and -- see~Eq.~\eqref{eq:NG-criterion} -- all of the non-Gaussian higher moments of the transverse momentum distribution $P({\bf k})$, including all of the transverse-longitudinal correlations,
are completely unsuppressed.
The upper bound in Eq.~\eqref{eq:bound}
informs us about the relative importance of non-Gaussian features of $P({\bf k})$ at the earliest times one could consider to make a measurement. In scenarios typical for heavy-ion collisions, the dynamics of heavy quarks in QGP will take place over a time interval that is a few times $1/T$, meaning that it is the upper bound that sets the characteristic size of non-Gaussianity. Non-Gaussian features may then be expected to be sizeable regardless, 
and they become dominant for sufficiently high velocity~\eqref{eq:NG-criterion-n},
provided that $\sqrt{\lambda} \sqrt{\gamma} \, T \ll M$. 
If, for example, we plug in numbers that are motivated by consideration of $b$-quarks in heavy ion collisions, say $M=20 T$ and $\lambda=12$, the speed limit~\eqref{eq:speed-limit} becomes $\gamma<33$ and the regime of validity~\eqref{eq:condition-Langevin} becomes $0.42\, \gamma^{1/4}<(T{\cal T})^{-1}<1$ or, equivalently with Eq.~\eqref{eq:bound}, $0.12\,\gamma^{3/4}<r<0.29\gamma^{1/2}$ which means that for $\gamma\sim 10$ and $T{\cal T}\sim 1$ the parameter that quantifies deviations from Gaussianity is $r\sim 0.92$ and non-Gaussian features of $P({\bf k})$ are very important indeed.
Our purpose in this paper is not phenomenology, but this estimate does suffice to show that the non-Gaussian features of $P({\bf k})$ that we have calculated for the first time must be taken into consideration in phenomenological studies. 
We comment that even if setting $\T = 1/T$ is at the edge of what our calculation allows, practical applications such as the one we will outline in the next Section are likely to use an even smaller value of $\T$, relying on the fact that a coarse-grained average over time scales of order $1/T$ will reproduce the same dynamics that we have calculated from first principles at longer time scales.

\section{Consequences for Heavy Quark Transport} \label{sec:consequences}

So far, our discussion has centered around the probability distribution $P({\bf k})$ defined by calculating the thermal average of the absolute value of the quantum mechanical scattering amplitude for a heavy quark with a velocity $v$ that satisfies the speed limit \eqref{eq:speed-limit} to change its momentum by ${\bf k}$ while propagating through a strongly coupled fluid, provided its velocity $v$ stays almost unchanged. In this Section, we extend our results by allowing $v$ to be a dynamical variable as well, to be identified with the total heavy quark momentum via ${\bf v} = {\bf p}/\sqrt{M^2 + {\bf p}^2}$, such that as the heavy quark transfers momentum to the QGP its $v$ decreases.
This will permit us to derive an evolution equation for the distribution of the total momentum of a heavy quark $\mathscr{P}({\bf p},t)$ as it propagates through the plasma, or as an ensemble of heavy quarks propagate through the plasma.\footnote{As noted in Ref.~\cite{Herzog:2006gh}, the mass $M$ used to convert between velocity and momentum, which is referred to as $M_{\rm kin}$ in that work, can be temperature-dependent and need not correspond to the rest mass of the heavy quark in vacuum.} Upon truncating this evolution equation for $\mathscr{P}$  to two derivatives, we shall recover an equation in the Fokker-Planck form which describes the evolution of an ensemble of heavy quarks each one of which moves according to a Langevin equation.  
Specifying this Fokker-Planck equation for the distribution $\mathscr{P}$ only requires knowledge of the 
mean and Gaussian variance of $P({\bf k})$, whereas the full evolution equation that we shall derive involves all the higher moments of $P({\bf k})$. 

To avoid later confusion, we emphasize that $P({\bf k},\T)$ and $\mathscr{P}({\bf p},t)$ represent different physical quantities.

\subsection{Kolmogorov and Fokker-Planck equations}

Our starting point in this Section is the same as the starting point that in a simplified context would lead to the Langevin equation. Given a heavy quark at position ${\bf x}$ and with momentum ${\bf p}$ at time $t$, one updates the state of the heavy quark at time $t + \Delta t$ via
\begin{align}
    {\bf x}(t + \Delta t) &= {\bf x}(t) + \Delta t \, {\bf v}(t) \label{eq:langevin-1} \, , \\
    {\bf p}(t + \Delta t) &= {\bf p}(t) - {\bf k}(t;\Delta t) \label{eq:langevin-2} \, .
\end{align}
Here, ${\bf k}(t;\Delta t)$ is to be sampled from the distribution $P({\bf k}; \T = \Delta t)$ that we have computed and discussed in the previous Section, and $t$ enters implicitly by defining ${\bf v}(t)$ in terms of the heavy quark momentum and $T(t)$ characterizing the background temperature. Crucially, because we have computed $P({\bf k})$ in full analytically, we need not make any assumptions about the ``random force'' encoded in ${\bf k}(t;\Delta t)$. (The usual assumption that takes one down a path to the Langevin equation is that there is a fluctuating force  described by a Gaussian-distributed random variable with a specified variance, or diffusion constant. See, for example, Ref.~\cite{Moore:2004tg}.)

In principle, Eqs.~\eqref{eq:langevin-1} and~\eqref{eq:langevin-2} fully specify how to propagate a heavy quark through a thermal plasma. However, in practice, 
one should first verify that after a total time $t_{\rm tot}$ the result of updating Eqs.~\eqref{eq:langevin-1} and~\eqref{eq:langevin-2} with $P({\bf k}; \T = \Delta t)$  is independent of the time step $\Delta t$. This is an important property, as it allows one to take $\Delta t < 1/T$, even if our previous calculation required $\T \gg 1/T$. It also guarantees that there is no ambiguity induced by different time discretizations of the stochastic 
equation~\eqref{eq:langevin-2}. 
The statement that the result of the stochastic process~\eqref{eq:langevin-1},~\eqref{eq:langevin-2} is independent of the choice of $\Delta t$ is equivalent to the statement that 
the addition property of independent random variables
\begin{equation}
    P({\bf k}; \T_1 + \T_2) = \int d^3 {\bf q} \, P({\bf k} - {\bf q} ; \T_1) P({\bf q} ; \T_2) \, \label{eq:addition-rv}
\end{equation}
holds identically for constant $v$ (i.e., in the $M \to \infty$ limit). This latter statement is non-trivial to check. If one starts from Eq.~\eqref{eq:Prob-result}, the composition property \eqref{eq:addition-rv} holds only to leading order in the strong coupling expansion. In contrast, if we focus on Eq.~\eqref{eq:P-from-S-first} which we restate here,
\begin{equation}
    P({\bf k}) = \frac{1}{(2\pi)^3} \int d^3 {\bf L} \, e^{ - i {\bf k} \cdot {\bf L} } \exp \left( - \sqrt{\lambda} T \T S_{\rm tot}({\bf L}) \right) \, , \label{eq:P-from-S}
\end{equation}
we see that it implies that~\eqref{eq:addition-rv} holds identically. Of course, Eqs.~\eqref{eq:Prob-result} and~\eqref{eq:P-from-S} are only
guaranteed to agree in the large $\lambda$ limit. We conclude that -- within the regime of validity of our calculation -- the dynamics of the momentum broadening is indeed independent of $\Delta t$. It is clear, however, that Eq.~\eqref{eq:P-from-S} is  more convenient for our present purposes.

Eq.~\eqref{eq:P-from-S} also serves to derive the concrete form of the generalized Fokker-Planck equation, or Kolmogorov equation, that $P({\bf k})$ satisfies. By differentiating \eqref{eq:P-from-S} with respect to $\T$ and then integrating by parts, one finds
\begin{equation}
    \partial_{\T} P = - \sqrt{\lambda} T S_{\rm tot}(  i \partial_{\bf k} ) P \, . \label{eq:FP}
\end{equation}
This equation updates  the momentum transfer probability distribution $P({\bf k})$ in time $\T$ provided that $v$ is held fixed. 
The moments of $P({\bf k})$ that follow from the expression~\eqref{eq:FP} are exactly those given by Eq.~\eqref{eq:moments-from-S}. And, truncating Eq.~\eqref{eq:FP} up to a finite number of derivatives with respect to ${\bf k}$ is equivalent to truncating $P({\bf k})$ such that only a finite number of its connected moments are nonzero.
As we explain now, the same equation~\eqref{eq:FP} can also be used to update the velocity of the heavy quark itself, and in this way to obtain an evolution equation for the probability distribution of the total momentum of the heavy quark $\mathscr{P}({\bf p},t)$.

To this end, we start from the probability of the heavy quark transitioning from a state with momentum ${\bf p}$ to a state with momentum ${\bf p}'$ over a time interval $\Delta t$, 
\begin{equation}
    \mathcal{P}({\bf p}'|{\bf p}; \Delta t) = P \left( {\bf k} = -({\bf p}' - {\bf p}) ; \, {\bf v} = \frac{\bf p}{\sqrt{{\bf p}^2 + M^2}} , \T = \Delta t \right) \, .
\end{equation}
Here, the minus sign in the ${\bf k}$ argument accounts for the fact that $P({\bf k})$ describes the transfer of momentum ${\bf k}$ from the heavy quark to the plasma.
From this expression, we see that the probability of finding a heavy quark with momentum ${\bf p}'$ at time $t + \Delta t$ satisfies 
\begin{equation}
    \mathscr{P}({\bf p}', t + \Delta t) = \int d^3 {\bf p} \, \mathcal{P}({\bf p}'|{\bf p}; \Delta t) \mathscr{P}({\bf p}, t) \, .
\end{equation}
We can now expand this expression for small $\Delta t$, integrate by parts, and proceed to derive an evolution equation for $\mathscr{P}({\bf p},t)$. The result is that the probability distribution for the heavy quark momentum (not just its change conditioned on a near-constant velocity $v$) is given by
\begin{equation}
    \partial_t \mathscr{P} = - \sqrt{\lambda} T K(\partial_{\bf p}, {\bf p}) \mathscr{P} 
    \label{eq:Kolmogorov} 
\end{equation}
where
\begin{equation}    
    K({\bf x}, {\bf p}) \equiv S_{\rm tot}\! \left( - i {\bf x} ; \, {\bf v} = \frac{{\bf p}}{\sqrt{{\bf p}^2 + M^2} } \right) \, . \label{eq:K-operator}
\end{equation}
In the expression \eqref{eq:K-operator} which defines the operator $K$, writing its first argument as $- i{\bf x}$ means that everywhere that ${\bf L}$ occurs in  $S_{\rm tot}({\bf L})$ it is to be replaced by $- i{\bf x} \rightarrow - i\partial_{\bf p}$.
In Eq.~\eqref{eq:Kolmogorov}, the derivatives in the first argument of $K$ act on its velocity argument as well as $\mathscr{P}$. That is to say, in a power series representation, these derivatives are inserted to the left of the rest of the expression in each term. We refer to this equation as a Kolmogorov equation, so as to distinguish it from its truncation, the Fokker-Planck equation which only involves derivatives up to second order. 
A Kolmogorov equation describes the evolution of a probability distribution in a Markov process (i.e., one whose update step only depends on the state of the system at the present time). The Fokker-Planck equation is a special, truncated, example.

In order to write Eq.~\eqref{eq:Kolmogorov} explicitly, we first write $S_{\rm tot}({\bf L})$ identifying $L_3 = \hat{p} \cdot {\bf L}$. 
Since ${\bf v}$ is now linked with ${\bf p}$, the generalized Fokker-Planck equation for $\mathscr{P}({\bf p},t)$ will exhibit explicit spherical symmetry. Restoring the explicit velocity dependence, we write $S_{\rm tot}({\bf L})$ in a power series and use the result
\eqref{eq:moments-from-S} to express the coefficients in terms of the connected moments of $P({\bf k})$ that we have calculated explicitly in Section~\ref{sec:non-gaussian}, obtaining
\begin{equation}
    S_{\rm tot} \! \left( {\bf L}; {\bf v} = \frac{{\bf p}}{\sqrt{{\bf p}^2 + M^2} } \right) = - \sum_{m,n=0}^{\infty} \frac{ i^{2m+n} }{(2m)! \, n!} \frac{\langle k_1^{2m} k_3^{n} \rangle_c }{\sqrt{\lambda} T \T} \left({\bf L}^2 - (\hat{p} \cdot {\bf L})^2 \right)^{m} \left(\hat{p} \cdot {\bf L} \right)^n \, .
\end{equation}
Note that the moments $\langle k_1^{2m} k_3^{n} \rangle_c$ depend on ${\bf p}$ through their velocity dependence. We then obtain the Kolmogorov equation for $\mathscr{P}({\bf p},t)$ in the form
\begin{equation}
    \partial_t \mathscr{P} = \sqrt{\lambda} T \sum_{m,n=0}^{\infty} \frac{1}{(2m)! \, n!} \left( \prod_{k=1}^m \partial_{i_k} \partial_{j_k} \right) \!\! \left( \prod_{k'=1}^n \partial_{l_{k'}} \right) \!\! \left[ \frac{\langle k_1^{2m} k_3^{n} \rangle_c }{\sqrt{\lambda} T \T} \prod_{k=1}^m \left( \delta_{i_k j_k} - \hat{p}_{i_k} \hat{p}_{j_k} \right) \prod_{k'=1}^n \hat{p}_{l_{k'}} \right] \mathscr{P} \, . \label{eq:FP-full}
\end{equation}
Here and below, the repeated indices $i_k$, $j_k$ and $l_{k'}$ are each summed from 1 to 3. Note also that in this expression all of the derivatives have been moved to the left, as discussed above.

Eq.~\eqref{eq:FP-full} uniquely and completely determines how the moments of the  distribution $P({\bf k})$, including all of the non-Gaussian moments, drive the dynamics of the momentum distribution of a heavy quark. In the previous Section, we have derived an explicit analytical recipe for calculating any and all of these non-Gaussian moments.
This means that in Eq.~\eqref{eq:FP-full}
we have now derived a complete description of how a heavy quark propagates through strongly coupled $\mathcal{N} = 4$ SYM plasma from first principles, taking every higher non-Gaussian moment of the fluctuations of its longitudinal and transverse momentum, and every correlation among these fluctuations, fully into account.

\subsection{The Equilibrium 
Distribution}
\label{sec:equil-distr}

We can now finally return to the thermalization problem posed in the Introduction and ask:  
Does the Kolmogorov evolution equation~\eqref{eq:FP-full} derived here admit a Boltzmann distribution $\mathscr{P} \propto \exp \left( - \sqrt{M^2 + {\bf p}^2 }/T \right)$ as a stationary solution?

The truncation of Eq.~\eqref{eq:FP-full} up to two derivatives is of the Fokker-Planck form,
\begin{align}
    \partial_t \mathscr{P} = \partial_i \! \left( \eta_D p_i \mathscr{P} \right) + \frac12  \partial_i \partial_j \! \left( \left[ \kappa_T \delta_{ij} + \left(\kappa_L - \kappa_T \right) \hat{p}_i \hat{p}_j \right]  \mathscr{P} \right) \, . \label{eq:FP-quad}
\end{align}
It is only this truncation of Eq.~\eqref{eq:FP-full} that had been considered in the literature~\cite{vanHees:2004gq,Beraudo:2009pe,He:2013zua}.
Using the classic results for $\eta_D, \kappa_T$ and $\kappa_L$ that we have reproduced in Section~\ref{sec:calculation}, Eq.~\eqref{eq:FP-quad} does \textit{not} have a Boltzmann distribution as a stationary solution, as was already highlighted by Gubser in the early work that we quoted in the Introduction. This has led many to dismiss these results in their application to phenomenology at $v>0$. Whenever some of these results have been used at $v>0$, at least one of them (typically $\kappa_L$) has been modified in some fashion so as to ensure that the distribution does reach thermal equilibrium, see for example 
Ref.~\cite{Beraudo:2009pe}. 

To find the stationary solution to
the complete evolution equation~\eqref{eq:FP-full} that $\mathscr{P}$ obeys, it is useful to introduce a rescaled momentum variable and to absorb the temperature dependence in rescaled moments: 
\begin{align}
    u_i \equiv \frac{p_i}{M} \, , & & c_{mn}(v) \equiv \frac{\langle k_1^{2m} k_3^{n} \rangle_c }{\sqrt{\lambda} T^{2m+n+1} \T} \, .
\end{align}
We note that $v = u/\sqrt{1+u^2}$, with $u \equiv |{\bf u}|$.  
With this notation, we can write
Eq.~\eqref{eq:FP-full} as 
\begin{equation}
    \partial_t \mathscr{P} = \sqrt{\lambda} T \sum_{m,n=0}^{\infty} \frac{(T/M)^{2m+n}}{(2m)! \, n!} \left( \prod_{k=1}^m \partial_{i_k} \partial_{j_k} \right) \!\! \left( \prod_{k'=1}^n \partial_{l_{k'}} \right) \!\! \left[ c_{mn}(v) \prod_{k=1}^m \left( \delta_{i_k j_k} - \hat{u}_{i_k} \hat{u}_{j_k} \right) \prod_{k'=1}^n \hat{u}_{l_{k'}} \right] \mathscr{P} \, , \label{eq:FP-full-u}
\end{equation}
where the derivatives are now with respect to $u_i$. 
In this way, we have organized the equation as a power series in $T/M$, a small parameter in our setup. This allows us to look for a stationary solution of the form
\begin{equation}
    \mathscr{P}(u) = \exp \left( - \sum_{a=-1}^\infty \left(\frac{T}{M}\right)^{a} f_a(u) \right) \, , \label{eq:FP-sol-form}
\end{equation}
where the sum over $a$ can be thought of as a perturbative expansion in $T/M$.

We emphasize that in Eq.~\eqref{eq:FP-sol-form}, the expansion in $T/M$ starts at power $a=-1$.\footnote{
The only stationary solution to Eq.~\eqref{eq:FP-full-u}
in which the series in Eq.~\eqref{eq:FP-sol-form} starts at $a=0$ is the solution  in which $\mathscr{P}$ is a $u$-independent constant.
} 
Physically, this is required of course for an ansatz that allows for an equilibrium distribution $\propto \exp(-E(p)/T)$. 
A nonzero coefficient for $a=-1$ 
has important consequences, 
as it implies that when any of the many derivatives with respect to $u$ in Eq.~\eqref{eq:FP-full-u} act on $\mathscr{P}$, they may pull down a power of $M/T$.
This in turn implies that even though in \eqref{eq:FP-sol-form} $\mathscr{P}(u)$ is written as a power series in $T/M$, the evolution equation \eqref{eq:FP-full-u} that determines the leading term $f_{-1}(u)$  
results from a competition between the first and all subsequent higher order terms in Eq.~\eqref{eq:FP-full-u}.
It is for this reason that truncating Eq.~\eqref{eq:FP-full-u} up to second derivatives (so as to obtain a Fokker-Planck equation) is not a consistent approximation. 
In contrast, finding any stationary solution to the full Kolmogorov evolution equation involves finding a balance between all the higher connected moments of $P({\bf k})$ which compete within \eqref{eq:FP-full-u} and contribute to its solution.

Formally, inserting the ansatz~\eqref{eq:FP-sol-form} into~\eqref{eq:FP-full-u} and requiring a stationary solution for each power of $T/M$, one finds an infinite hierarchy of equations for all $f_a(u)$, $a=-1,0,1,2,\ldots$. 
The first equation in this hierarchy is for $f_{-1}(u)$. With the results obtained in this work we can only give the final answer for $f_{-1}$, because any of the rescaled moments $c_{mn}$ (all of which contribute, as noted above) may have corrections controlled by $T/M$, 
which are outside the scope of our calculation. Such corrections would modify the solutions for $f_a$ with $a \geq 0$ relative to what we could obtain from~\eqref{eq:FP-full-u}.  

For the coefficient $f_{-1}(u)$, inserting~\eqref{eq:FP-sol-form} into~\eqref{eq:FP-full-u} and demanding that the solution be stationary yields
\begin{equation}
    \sum_{n=1}^\infty \frac{(-1)^n}{n!} c_{0n}(v) \big( f_{-1}'(u) \big)^n = S_{\rm tot}(L_\perp = 0, L_3 = i f_{-1}'(u)/T) = 0 \, , \label{eq:condition-to-solve-equil-FP}
\end{equation}
where $v = u/\sqrt{1+u^2}$. It is important to note that, from the point of view of Eq.~\eqref{eq:FP-full-u}, this equation involves terms of arbitrary high order in $T/M$. As mentioned above, this is a direct consequence of the fact that the perturbative expansion in Eq.~\eqref{eq:FP-sol-form} starts at power $a=-1$. Equation \eqref{eq:condition-to-solve-equil-FP} makes explicit the resulting competition between the leading and {\it all} higher orders, which are characterized by the non-Gaussian moments we have calculated. It is then obvious that the truncation of Eq.~\eqref{eq:FP-sol-form} up to second order in derivatives will not have the same stationary solution.

Using Eqs.~\eqref{eq:L-C-relation} and~\eqref{eq:S-Stot-relation}, we now show how to find  $f'_{-1}$ explicitly. For brevity, we define $x \equiv -f_{-1}'(u)/T$. To find a solution to Eq.~\eqref{eq:condition-to-solve-equil-FP}, we need to solve
\begin{equation}
    \left[\tilde{S}_{\rm tot} - C_3 \frac{\partial \tilde{S}_{\rm tot} }{\partial C_3} \right]_{C_3 = \bar{C}_3(x), C_\perp = 0} = 0, \label{eq:condition-to-solve-equil-FP-2}
\end{equation}
where $\bar{C}_3(x)$ is determined by
\begin{equation}
    x = 2 \left. \frac{\partial \tilde{S}_{\rm tot} }{\partial C_3} \right|_{C_3 = \bar{C}_3(x), C_\perp = 0} \, . \label{eq:condition-to-solve-equil-FP-3}
\end{equation}
It is easy to check that there are only two values of $C_3$ for which Eq.~\eqref{eq:condition-to-solve-equil-FP-2} holds, determined by the configurations where $z_+ = z_-$. The solutions are $C_3 = \pm \pi T v \gamma$, where $v > 0$. The corresponding values of $x$ are given by Eq.~\eqref{eq:condition-to-solve-equil-FP-3}. One of them is $x = 0$, determined by $\bar{C}_3(x) = \pi T v \gamma$, corresponding to a trivial solution to the Kolmogorov equation $f_{-1}'=0$. The other is $x = -v/T$, determined by $\bar{C}_3(x) = - \pi T v \gamma$ [note that at this point $\partial \tilde{S}_{\rm tot}/\partial C_3 = - v/(2T)$ is not zero only because of the term proportional to the Heaviside step function in Eq~\eqref{eq:stot-result}], which is a nontrivial solution, as it determines
\begin{equation}
    f_{-1}'(u) = v = \frac{u}{\sqrt{1+u^2}} \implies f_{-1}(u) = \sqrt{1 + u^2} \, ,
\end{equation}
which then implies that
\begin{equation}
    \mathscr{P} = \exp \left( - \frac{\sqrt{M^2 + {\bf p}^2 }}{T} - \sum_{a=0}^\infty \left(\frac{T}{M}\right)^{a} f_a \right) \, .
    \label{boltzmann}
\end{equation}
Therefore, up to the order we can reliably calculate in an expansion in $T/M$, heavy quarks do thermalize precisely as one would expect from general principles in statistical physics. We expect that this conclusion will be unchanged if one calculates the moments $\langle k_1^{2m} k_3^n \rangle_c$ up to the necessary order in $T/M$ to fully determine the subleading $f_a$ functions. This solves the longstanding problem that we had recalled in the Introduction, namely that previous analyses of heavy quark propagation in strongly coupled $\mathcal{N}=4$ plasma had not been able to show that the dynamics could drive their momentum distribution to local thermal equilibrium. We have just done exactly that.

We emphasize that this result is noteworthy for several additonal reasons:
\begin{enumerate}
    \item The fact that the stationary solution is a Boltzmann distribution is a consequence of a precise balancing between \textit{all} terms in the series expansion that defines the Kolmogorov equation~\eqref{eq:FP-full}, involving arbitrary high non-Gaussian moments of $P({\bf k})$, not just the terms with up to two derivatives as in a Fokker-Planck equation. 
    
    \item The system thermalizes despite the fact that there is no fluctuation-dissipation relation between the coefficients of the first derivatives and the second derivatives in~\eqref{eq:FP-full}.
    \item Finding a nontrivial solution to Eq.~\eqref{eq:condition-to-solve-equil-FP} required us to know the precise form of the term that breaks the $C_3 \to -C_3$ symmetry in Eq.~\eqref{eq:stot-result}. Conversely, if nothing in our formulas guaranteed that energy loss is likelier than energy gain, such an equilibrium solution would not exist.
    \item After completing the calculation that we have described, one can ask whether the form of the dispersion relation for the heavy quark is important to the conclusion that the Kolmogorov equation does indeed describe thermalization. If one goes back through the derivation, one finds that the precise form of the dispersion relation of the heavy quark $E(p) = \sqrt{M^2 + {\bf p}^2 }$ is in fact irrelevant to the form of the solution to~\eqref{eq:FP-full} (provided, of course, that there exists a large scale $M$ to justify our approximations). All of the moments $\langle k_1^{2m} k_3^n \rangle_c$ that we have calculated as functions of $v$ are important, as we have discussed, but the solution to Eq.~\eqref{eq:condition-to-solve-equil-FP} will still be given by $f_{-1}' = v$, regardless of the details of the dispersion relation, since for any dispersion relation $E(p)$ the group velocity $v$ and the dispersion relation are related by
    \begin{equation}
        v = \frac{\partial E}{\partial p}\ .
    \end{equation}
    This motivates further investigation of just how general the remarkable conclusions that we have found are. A natural starting point for such an investigation will be to consider other (possibly nonconformal) theories with dual gravitational descriptions.

\end{enumerate}

Having succeeded in fully
resolving the puzzle first articulated more than fifteen years ago that we began with in the Introduction, for the first time we have a complete and consistent description of heavy quark transport and thermalization in strongly coupled ${\cal N}=4$ SYM plasma.

\section{Concluding Remarks and a Look Ahead} 
\label{sec:conclusions}

In Sections~\ref{sec:ads-cft} and \ref{sec:calculation} of the present work, we have formulated and implemented a calculational 
setup that provides a unified description of all the Gaussian characteristics of the momentum transfer probability distribution $P({\bf k})$ for heavy quarks propagating through the strongly coupled plasma of ${\cal N}=4$ SYM theory
that have been calculated previously,
namely 
the drag coefficient $\eta_D$, the transverse and longitudinal momentum diffusion coefficients $\kappa_T$ and $\kappa_L$, and the jet quenching parameter $\hat{q}$. Beyond placing  all these previous results in a common framework, our calculation of $P({\bf k})$ has allowed us, in 
Section~\ref{sec:non-gaussian},
to determine all the higher non-Gaussian moments 
of $P({\bf k})$ including those that describe correlations between transverse and longitudinal momentum fluctuations completely, for the first time. 
As an illustrative example of the importance of higher moments to phenomenology, we can make an estimate that indicates that the skewness of the longitudinal momentum fluctuations is large enough to significantly reduce the stopping length for heavy quarks.
With  $P({\bf k})$ in hand, in 
Section~\ref{sec:consequences} we have derived the Kolmogorov evolution equation~\eqref{eq:FP-full} for heavy quark transport which knows about all the Gaussian and non-Gaussian moments 
of $P({\bf k})$ that we have calculated explicitly in Section~\ref{sec:moments}. If truncated at second order in derivatives, this Kolmogorov equation~\eqref{eq:FP-full} reduces to the known Fokker-Planck equation in terms of $\eta_D$, $\kappa_T$ and $\kappa_L$, only. This truncated evolution does not allow for the Boltzmann distribution as a stationary solution and it thus exhibits the long-standing thermalization problem that we have recalled in the Introduction. In contrast, we have shown explicitly in Section~\ref{sec:consequences} that the stationary equilibrium solution of the complete untruncated evolution equation~\eqref{eq:FP-full}  {\it is} the Boltzmann distribution. 
Remarkably, this only arises as a consequence of a balance among competing contributions from all the higher moments of $P({\bf k})$ in the full evolution equation~\eqref{eq:FP-full} at leading order in $T/M$.
That is, the non-Gaussian moments determined here for the first time are essential for ensuring 
that heavy quark transport describes the approach to equilibrium expected on physical grounds.

The next step, which we leave to future work,  is to study Eq.~\eqref{eq:FP-full} away from equilibrium. For the out-of-equilibrium dynamics of $f_{-1}$, which is the leading contribution in $T/M$ to the distribution function $\mathscr{P}$, the Kolmogorov equation reduces to 
\begin{equation}
    \frac{M}{\sqrt{\lambda} T^2} \frac{\partial f_{-1}}{\partial t}  =  K \! \left( - \frac{M}{T} \frac{\partial f_{-1}}{\partial {\bf p}} , {\bf p} \right) \, , \label{eq:f-1-out-of-eq}
\end{equation}
where $K$ is given in Eq.~\eqref{eq:K-operator}, and where we have restored all dimensionful quantities. Therefore, if we only keep this contribution to $\mathscr{P}$, we have
\begin{equation}
    \frac{1}{\sqrt{\lambda} T} \frac{\partial \ln \mathscr{P} }{\partial t} = - K \! \left( \frac{\partial \ln \mathscr{P}}{\partial {\bf p}} , {\bf p} \right) \, . \label{eq:P-out-of-eq-1st}
\end{equation}
Without providing further results in the present work, let us mention that our preliminary numerical studies indicate that perturbations evolved with Eq.~\eqref{eq:P-out-of-eq-1st} do indeed evolve to the stationary equilibrium Boltzmann distribution. We thus have the first evidence for our physical expectation that, at leading order in $T/M$, the stationary solution \eqref{boltzmann} of the evolution equation \eqref{eq:P-out-of-eq-1st} is a universal attractor solution to which arbitrary out-of-equilibrium distributions evolve. We emphasize that the derivation of Eq.~\eqref{eq:P-out-of-eq-1st} is not limited to small perturbations: this evolution equation provides a dynamically complete framework for understanding the thermalization of heavy quarks that may be initially plowing through the plasma at a high speed, far from equilibrium. 

Starting from the present work, we see numerous directions for further investigation. One direction is clearly to base studies of out-of-equilibrium evolution on Eq.~\eqref{eq:FP-full}. At first glance from Eq.~\eqref{eq:f-1-out-of-eq}, the (inverse) time scale of equilibration is given by $\sqrt{\lambda} T^2/M \propto \eta_D$. However, the explicit velocity/momentum dependence of $K$ will certainly give rise to a different rate of approach to equilibrium for different values of ${\bf p}$. The determination of a stopping distance and other characteristics of the process of equilibration 
will involve more than just 
the single relaxation parameter $\eta_D$. Given that the existence of a stationary equilibrium state depends on all the higher moments of $P({\bf k})$, we may expect that they all play a role in the determination of the stopping length also.
We shall leave a comprehensive phenomenological investigation of quantities like the stopping length, starting from the evolution equation~\eqref{eq:FP-full}, to future work.

We can, however, make an initial estimate of how significant
the modification of
the stopping length of a heavy quark arising from non-Gaussian fluctuations in the longitudinal momentum is likely to be.
An estimate may be obtained by correcting the mean momentum change $\langle k_3 \rangle$ we calculated (which corresponds to the value of the maximum of $P({\bf k})$, but does not determine the mean beyond the leading term in $\sqrt{\lambda} \T$) using the skewness of the longitudinal momentum fluctuations that we have calculated explicitly in Eq.~\eqref{eq:skewness}.  As we noted there, this positive skewness means that fluctuations toward much larger energy loss are more common than fluctuations toward much smaller energy loss. This should reduce the stopping length. With this motivation, we estimate
    \begin{equation}
        \Delta p_L \approx - \left[ \langle k_3 \rangle + \langle (k_3 - \langle k_3 \rangle )^3 \rangle^{1/3} \right] \, , \label{eq:pL-drag-shift}
    \end{equation}
    where the first term $\propto \eta_D$ encodes the long-known drag force~\cite{Gubser:2006bz,Herzog:2006gh} and is the leading contribution in the $\lambda\to\infty$ limit while
    the second term, for which we now have an explicit form in Eq.~\eqref{eq:skewness}, takes account of the shift to the mean of the distribution due to the skewness (the first non-Gaussian moment) of the longitudinal momentum fluctuations.  Taking a time step of $\T = \Delta t = 1/T$, we may convert Eq.~\eqref{eq:pL-drag-shift} into a differential equation as a function of the distance traveled $x$, 
    \begin{equation}
        \frac{dE}{dx} = - \eta_D \, E\left[ \frac{\sqrt{E^2 - M^2}}{E} + \left(\frac{6}{\pi \sqrt{\lambda}}\right)^{2/3} \frac{(E^2-M^2)^{1/6}}{M^{1/3}} \right] \, ,
        \label{eq:stopping-length-equation}
    \end{equation}
    which 
   upon integration from $E=E_{\rm init}$ to $E=M$  
    determines the stopping length of a heavy quark, as first discussed (with only the leading term on the right-hand-side) in Ref.~\cite{Herzog:2006gh}. 
Noting that $\eta_D\propto \sqrt{\lambda}$, we see that the contribution to \eqref{eq:stopping-length-equation} arising from the skewness of the longitudinal momentum fluctuations is parametrically of order $\lambda^{1/6}$. This means that, although it is suppressed relative to the leading term, it is still parametrically larger than the subleading contributions in the saddle point approximation that we have used, which start at order $\lambda^0$.
Although the second term is formally subleading, it is easy to see that it can be quite significant. For example, if we take $\lambda=12$ the second term on the right-hand-side of Eq.~\eqref{eq:stopping-length-equation} is 1.17 times larger (1.45 times larger) than the leading term for heavy quarks with $E=5M$ ($E=10M$). Or, if we take $M=20 T$ (motivated by $b$-quarks in the QGP produced in a heavy ion collision) and $\lambda=12$ and evaluate the stopping length for a heavy quark with $E_{\rm init}=5 M$ by integrating Eq.~\eqref{eq:stopping-length-equation} we find a stopping length of $8.42/T$ if we only keep the leading term in Eq.~\eqref{eq:stopping-length-equation} while including the skewness term in addition yields a stopping length of $3.90/T$. 
These numbers should not be taken too seriously because the fact that the modification to the stopping length originating from the skewness of the longitudinal momentum fluctuations is so significant means that contributions coming from even higher moments, which also contribute positively to $dE/dx$, are likely significant also.  This motivates the phenomenological importance of a future investigation starting from the full evolution equation~\eqref{eq:FP-full}.

Another promising direction for future work would be to generalize Eq.~\eqref{eq:P-out-of-eq-1st} to a hydrodynamic background, where both the velocity and the temperature of the plasma are functions of space and time (and therefore no global rest frame exists). We expect that the non-Gaussian moments we have calculated in this work will play a crucial role in such a generalization.

More generally, one can ask what insights from our results, obtained via gauge-gravity duality for the strongly coupled ${\cal N}=4$ SYM plasma, 
are likely to be relevant for other strongly coupled non-abelian plasmas including the QGP formed in heavy-ion collisions. 
We expect that non-Gaussian higher moments of $P({\bf k})$ are nonzero, and it is likely that they become more significant with increasing $\gamma$. 
We do not expect that the 
precise values of the moments $\langle k_1^{2m} k_3^{n} \rangle_c$ of $P({\bf k})$ 
are the same in strongly coupled QGP as those we have
calculated in ${\cal N}=4$ SYM, but it will be very interesting to investigate whether the way they scale with $\gamma$ generalizes.
And,
there is no reason at all to expect $P({\bf k_\perp},k_L)$ to be symmetric in $k_L-\langle k_L\rangle$, and it seems likely to be generic for the asymmetry introduced by the non-Gaussian moments to favor more-than-mean energy loss over less-than-mean energy loss.
Once we have seen in a completely rigorous fashion that in ${\cal N}=4$ SYM
theory the Einstein relation
\eqref{eq:Einstein} that is familiar and correct at $v=0$ 
breaks down
as the non-Gaussianity of $P({\bf k})$ develops with increasing $\gamma$, and that the Fokker-Planck description that relies solely on Gaussian characteristics of $P({\bf k})$ becomes inadequate, one realizes that there is no reason at all to expect anything different in QCD. 
We therefore expect that
the solution to the thermalization problem 
for heavy quarks in QGP in QCD
cannot be found by adjusting the values of the Gaussian diffusion coefficients at nonzero $\gamma$ so as to impose the Einstein relation by hand. Rather, it lies in embracing the increasingly non-Gaussian shape of $P({\bf k})$ with increasing $\gamma$ and incorporating this information in a more complete evolution equation~\eqref{eq:FP-full} that goes beyond the Fokker-Planck truncation. Earlier indications for this have been seen in Relativistic Boltzmann Transport studies~\cite{Das:2013kea}.

We close by looking ahead to the experimental and theoretical challenges in heavy ion phenomenology; for a review, see Ref.~\cite{Busza:2018rrf}. Multiple measurements, including measurements of the nuclear suppression factor $R_{AA}$ and of the anisotropic elliptic flow $v_2$ of heavy-flavored hadrons (via their hadronic or semi-leptonic decay products) at RHIC and at the LHC provide by now a consistent picture of heavy flavor transport in the QGP produced in these collisions, see Ref.~\cite{Dong:2019byy} for a review. Improving the accuracy of these measurements and achieving a precise determination of the heavy quark diffusion coefficient is one of the main motivations for experimental upgrades and for a new detector that could be operational in LHC run 5~\cite{ ALICE:2022wwr}. These ongoing experimental efforts are paralleled by more than two decades of theoretical work towards a consistent and phenomenologically satisfactory description of heavy flavor transport in the QGP produced in nucleus-nucleus collisions. The current state of the art is based on many important works including Refs.~\cite{Moore:2004tg,vanHees:2005wb,Alberico:2011zy,Alberico:2013bza,Cao:2013ita} and has been reviewed in Refs.~\cite{Prino:2016cni,Cao:2018ews,Dong:2019unq,He:2022ywp}. The significant remaining model-dependent differences have been the subject of several working group 
reports~\cite{Xu:2018gux,Rapp:2018qla}. Without reviewing the current phenomenological practice in detail, let us note that the main body of model studies employ Langevin-type descriptions of heavy quark transport or, equivalently, Boltzmann transport formulations in the Fokker-Planck approximation. The Einstein relation \eqref{eq:Einstein} is then typically enforced by defining the longitudinal diffusion coefficient in terms of the drag coefficient. While this ad hoc choice provides a physically motivated remedy for realizing an equilibrating dynamics {\it within} a Langevin or Fokker-Planck formulation, our work indicates that a more thorough treatment of heavy quark transport needs to go {\it beyond} a Gaussian characterization of heavy quark diffusion. Phenomenologically, this is of interest not only because we have learned in the present work why the parametric $\gamma$-dependence of $\kappa_L$ need not be constrained by $\eta_D$, but also because such a more encompassing formulation of heavy quark transport may provide experimental access to higher moments $\langle k_1^{2m} k_3^{n} \rangle_c$ that determine cross-correlations between the longitudinal and transverse dynamics, that can be predicted in quantum field theory and that have not been explored so far. For all these reasons, we expect that further exploration of the Kolmogorov equation~\eqref{eq:FP-full} will provide a solid foundation for future phenomenological studies of heavy quark transport in non-abelian plasmas and may reveal profound features of QCD.

\acknowledgments

We thank Carlota Andres, Xi Dong, Jean Du Plessis, Giuliano Giacalone, Vincenzo Greco, Arjun Kudinoor, Hong Liu, Robinson Mancilla, David Mateos, Govert Nijs, Dani Pablos, Ralf Rapp, Andrey Sadofyev, Kostas Skenderis, Rachel Steinhorst, Balt van Rees and Xiaojun Yao for useful discussions. KR is grateful to the CERN Theory Department for hospitality and support. This research was supported in part by the U.S.~Department of Energy, Office of Science, Office of Nuclear Physics under grant Contract Number DE-SC0011090\@. The research of BSH was supported in part by grant NSF PHY-2309135 to the Kavli Institute for Theoretical Physics (KITP). The research of BSH was supported in part by grant 994312 from the Simons Foundation.

\bibliography{main.bib}

\end{document}